\documentstyle[12pt]{article}

\newtheorem{theorem}{Theorem}
\newtheorem{corollary}[theorem]{Corollary}
\newtheorem{conjecture}[theorem]{Conjecture}
\newtheorem{lemma}[theorem]{Lemma}
\newtheorem{proposition}[theorem]{Proposition}
\newtheorem{definition}[theorem]{Definition}
\newtheorem{example}[theorem]{Example}
\newtheorem{axiom}{Axiom}
\newtheorem{remark}{Remark}
\newtheorem{exercise}{Exercise}[section]

\makeatletter
\newcount\@hour\newcount\@minute\chardef\@x10\chardef\@xv60
\def\tcitime{
\def\@time{%
  \@minute\time\@hour\@minute\divide\@hour\@xv
  \ifnum\@hour<\@x 0\fi\the\@hour:%
  \multiply\@hour\@xv\advance\@minute-\@hour
  \ifnum\@minute<\@x 0\fi\the\@minute
  }}%
\def\QCTOpt[#1]#2{%
  \def\QCTOptB{#1}
  \def\QCTOptA{#2}
}
\def\QCTNOpt#1{%
  \def\QCTOptA{#1}
  \let\QCTOptB\empty
}
\def\Qct{%
  \@ifnextchar[{%
    \QCTOpt}{\QCTNOpt}
}
\def\QCBOpt[#1]#2{%
  \def\QCBOptB{#1}
  \def\QCBOptA{#2}
}
\def\QCBNOpt#1{%
  \def\QCBOptA{#1}
  \let\QCBOptB\empty
}
\def\Qcb{%
  \@ifnextchar[{%
    \QCBOpt}{\QCBNOpt}
}
\def\PrepCapArgs{%
  \ifx\QCBOptA\empty
    \ifx\QCTOptA\empty
      {}%
    \else
      \ifx\QCTOptB\empty
        {\QCTOptA}%
      \else
        [\QCTOptB]{\QCTOptA}%
      \fi
    \fi
  \else
    \ifx\QCBOptA\empty
      {}%
    \else
      \ifx\QCBOptB\empty
        {\QCBOptA}%
      \else
        [\QCBOptB]{\QCBOptA}%
      \fi
    \fi
  \fi
}
\newcount\GRAPHICSTYPE
\GRAPHICSTYPE=\z@
\def\GRAPHICSPS#1{%
 \ifcase\GRAPHICSTYPE
   \special{ps: #1}%
 \or
   \special{language "PS", include "#1"}%
 \fi
}%
\def\graffile#1#2#3#4{%
    \leavevmode
    \raise -#4 \BOXTHEFRAME{%
        \hbox to #2{\raise #3\hbox{\null #1}}}%
}%
\def\draftbox#1#2#3#4{%
 \leavevmode\raise -#4 \hbox{%
  \frame{\rlap{\protect\tiny #1}\hbox to #2%
   {\vrule height#3 width\z@ depth\z@\hfil}%
  }%
 }%
}%
\newcount\draft
\draft=\z@

\newif\ifwasdraft
\wasdraftfalse

\def\GRAPHIC#1#2#3#4#5{%
 \ifnum\draft=\@ne\draftbox{#2}{#3}{#4}{#5}%
  \else\graffile{#1}{#3}{#4}{#5}%
  \fi
 }%
\def\addtoLaTeXparams#1{%
    \edef\LaTeXparams{\LaTeXparams #1}}%
\newif\ifBoxFrame \BoxFramefalse
\newif\ifOverFrame \OverFramefalse
\newif\ifUnderFrame \UnderFramefalse

\def\BOXTHEFRAME#1{%
   \hbox{%
      \ifBoxFrame
         \frame{#1}%
      \else
         {#1}%
      \fi
   }%
}

\def\doFRAMEparams#1{\BoxFramefalse\OverFramefalse\UnderFramefalse\readFRAMEparams#1\end}%
\def\readFRAMEparams#1{%
 \ifx#1\end%
  \let\next=\relax
  \else
  \ifx#1i\dispkind=\z@\fi
  \ifx#1d\dispkind=\@ne\fi
  \ifx#1f\dispkind=\tw@\fi
  \ifx#1t\addtoLaTeXparams{t}\fi
  \ifx#1b\addtoLaTeXparams{b}\fi
  \ifx#1p\addtoLaTeXparams{p}\fi
  \ifx#1h\addtoLaTeXparams{h}\fi
  \ifx#1X\BoxFrametrue\fi
  \ifx#1O\OverFrametrue\fi
  \ifx#1U\UnderFrametrue\fi
  \ifx#1w
    \ifnum\draft=1\wasdrafttrue\else\wasdraftfalse\fi
    \draft=\@ne
  \fi
  \let\next=\readFRAMEparams
  \fi
 \next
 }%
\def\IFRAME#1#2#3#4#5#6{%
      \bgroup
      \let\QCTOptA\empty
      \let\QCTOptB\empty
      \let\QCBOptA\empty
      \let\QCBOptB\empty
      #6%
      \parindent=0pt%
      \leftskip=0pt
      \rightskip=0pt
      \setbox0 = \hbox{\QCBOptA}%
      \@tempdima = #1\relax
      \ifOverFrame
          \typeout{This is not implemented yet}%
          \show\HELP
      \else
         \ifdim\wd0>\@tempdima
            \advance\@tempdima by \@tempdima
            \ifdim\wd0 >\@tempdima
               \textwidth=\@tempdima
               \setbox1 =\vbox{%
                  \noindent\hbox to \@tempdima{\hfill\GRAPHIC{#5}{#4}{#1}{#2}{#3}\hfill}\\%
                  \noindent\hbox to \@tempdima{\parbox[b]{\@tempdima}{\QCBOptA}}%
               }%
               \wd1=\@tempdima
            \else
               \textwidth=\wd0
               \setbox1 =\vbox{%
                 \noindent\hbox to \wd0{\hfill\GRAPHIC{#5}{#4}{#1}{#2}{#3}\hfill}\\%
                 \noindent\hbox{\QCBOptA}%
               }%
               \wd1=\wd0
            \fi
         \else
            \ifdim\wd0>0pt
              \hsize=\@tempdima
              \setbox1 =\vbox{%
                \unskip\GRAPHIC{#5}{#4}{#1}{#2}{0pt}%
                \break
                \unskip\hbox to \@tempdima{\hfill \QCBOptA\hfill}%
              }%
              \wd1=\@tempdima
           \else
              \hsize=\@tempdima
              \setbox1 =\vbox{%
                \unskip\GRAPHIC{#5}{#4}{#1}{#2}{0pt}%
              }%
              \wd1=\@tempdima
           \fi
         \fi
         \@tempdimb=\ht1
         \advance\@tempdimb by \dp1
         \advance\@tempdimb by -#2%
         \advance\@tempdimb by #3%
         \leavevmode
         \raise -\@tempdimb \hbox{\box1}%
      \fi
      \egroup%
}%
\def\DFRAME#1#2#3#4#5{%
 \begin{center}
     \let\QCTOptA\empty
     \let\QCTOptB\empty
     \let\QCBOptA\empty
     \let\QCBOptB\empty
     \ifOverFrame 
        #5\QCTOptA\par
     \fi
     \GRAPHIC{#4}{#3}{#1}{#2}{\z@}
     \ifUnderFrame 
        \par #5\QCBOptA
     \fi
 \end{center}%
 }%
\def\FFRAME#1#2#3#4#5#6#7{%
 \begin{figure}[#1]%
  \let\QCTOptA\empty
  \let\QCTOptB\empty
  \let\QCBOptA\empty
  \let\QCBOptB\empty
  \ifOverFrame
    #4
    \ifx\QCTOptA\empty
    \else
      \ifx\QCTOptB\empty
        \caption{\QCTOptA}%
      \else
        \caption[\QCTOptB]{\QCTOptA}%
      \fi
    \fi
    \ifUnderFrame\else
      \label{#5}%
    \fi
  \else
    \UnderFrametrue%
  \fi
  \begin{center}\GRAPHIC{#7}{#6}{#2}{#3}{\z@}\end{center}%
  \ifUnderFrame
    #4
    \ifx\QCBOptA\empty
      \caption{}%
    \else
      \ifx\QCBOptB\empty
        \caption{\QCBOptA}%
      \else
        \caption[\QCBOptB]{\QCBOptA}%
      \fi
    \fi
    \label{#5}%
  \fi
  \end{figure}%
 }%
\newcount\dispkind%
\def\FRAME#1#2#3#4#5#6#7#8{%
 \ifnum\draft=\@ne
   \wasdrafttrue
 \else
   \wasdraftfalse%
 \fi
 \def\LaTeXparams{}%
 \dispkind=\z@
 \def\LaTeXparams{}%
 \doFRAMEparams{#1}%
 \ifnum\dispkind=\z@\IFRAME{#2}{#3}{#4}{#7}{#8}{#5}\else
  \ifnum\dispkind=\@ne\DFRAME{#2}{#3}{#7}{#8}{#5}\else
   \ifnum\dispkind=\tw@
    \edef\@tempa{\noexpand\FFRAME{\LaTeXparams}}%
    \@tempa{#2}{#3}{#5}{#6}{#7}{#8}%
    \fi
   \fi
  \fi
  \ifwasdraft\draft=1\else\draft=0\fi{}%
 }%
\def\TEXUX#1{"texux"}

\long\def\QQQ#1#2{%
     \long\expandafter\def\csname#1\endcsname{#2}}%
\@ifundefined{QTP}{\def\QTP#1{}}{}
\@ifundefined{Qlb}{}{}
\@ifundefined{Qlt}{}{}
\long\def\QQA#1#2{}%
\def\QTR#1#2{{\csname#1\endcsname #2}}
\long\def\TeXButton#1#2{#2}%
\def\EXPAND#1[#2]#3{}%
\def\NOEXPAND#1[#2]#3{}%
\def\LaTeXparent#1{}%
\def\ChildStyles#1{}%
\def\ChildDefaults#1{}%
\def\QTagDef#1#2#3{}%
\def\QQfnmark#1{\footnotemark}

\@ifundefined{INDEX}{\def\INDEX#1#2{}{}}{}%
\@ifundefined{SUBINDEX}{\def\SUBINDEX#1#2#3{}{}{}}{}%
\def\initial#1{\bigbreak{\raggedright\large\bf #1}\kern 2\p@
   \penalty3000}%
\@ifundefined{ZZZ}{}{\makeatletter\input gnuindex.sty\makeatother\makeindex\makeatletter}%
\@ifundefined{abstract}{%
 \def\abstract{%
  \if@twocolumn
   \section*{Abstract (Not appropriate in this style!)}%
   \else \small 
   \begin{center}{\bf Abstract\vspace{-.5em}\vspace{\z@}}\end{center}%
   \quotation 
   \fi
  }%
 }{%
 }%
\@ifundefined{endabstract}{\def\endabstract
  {\if@twocolumn\else\endquotation\fi}}{}%
\@ifundefined{maketitle}{\def\maketitle#1{}}{}%
\@ifundefined{affiliation}{\def\affiliation#1{}}{}%
\@ifundefined{proof}{\def\proof{\noindent{\bfseries Proof. }}}{}%
\@ifundefined{endproof}{\def\endproof{\mbox{\ \rule{.1in}{.1in}}}}{}%
\@ifundefined{newfield}{\def\newfield#1#2{}}{}%
\@ifundefined{chapter}{\def\chapter#1{\par(Chapter head:)#1\par }%
 \newcount\c@chapter}{}%
\@ifundefined{part}{\def\part#1{\par(Part head:)#1\par }}{}%
\@ifundefined{section}{\def\section#1{\par(Section head:)#1\par }}{}%
\@ifundefined{subsection}{\def\subsection#1%
 {\par(Subsection head:)#1\par }}{}%
\@ifundefined{subsubsection}{\def\subsubsection#1%
 {\par(Subsubsection head:)#1\par }}{}%
\@ifundefined{paragraph}{\def\paragraph#1%
 {\par(Subsubsubsection head:)#1\par }}{}%
\@ifundefined{subparagraph}{\def\subparagraph#1%
 {\par(Subsubsubsubsection head:)#1\par }}{}%
\@ifundefined{therefore}{}{}%
\@ifundefined{backepsilon}{}{}%
\@ifundefined{yen}{\def\yen{\hbox{\rm\rlap=Y}}}{}%
\@ifundefined{registered}{%
   \def\registered{\relax\ifmmode{}\r@gistered
                    \else$\m@th\r@gistered$\fi}%
 \def\r@gistered{^{\ooalign
  {\hfil\raise.07ex\hbox{$\scriptstyle\rm\text{R}$}\hfil\crcr
  \mathhexbox20D}}}}{}%
\@ifundefined{Eth}{}{}%
\@ifundefined{eth}{}{}%
\@ifundefined{Thorn}{}{}%
\@ifundefined{thorn}{}{}%
\@ifundefined{degree}{}{}%
\newdimen\theight
\def\Column{%
 \vadjust{\setbox\z@=\hbox{\scriptsize\quad\quad tcol}%
  \theight=\ht\z@\advance\theight by \dp\z@\advance\theight by \lineskip
  \kern -\theight \vbox to \theight{%
   \rightline{\rlap{\box\z@}}%
   \vss
   }%
  }%
 }%
\def\qed{%
 \ifhmode\unskip\nobreak\fi\ifmmode\ifinner\else\hskip5\p@\fi\fi
 \hbox{\hskip5\p@\vrule width4\p@ height6\p@ depth1.5\p@\hskip\p@}%
 }%
\def\miss{\hbox{\vrule height2\p@ width 2\p@ depth\z@}}%
\def\tcol#1{{\baselineskip=6\p@ \vcenter{#1}} \Column}  %
\def\newfmtname{LaTeX2e}
\def\chkcompat{%
   \if@compatibility
   \else
     \usepackage{latexsym}
   \fi
}

\ifx\fmtname\newfmtname
  \DeclareOldFontCommand{\rm}{\normalfont\rmfamily}{\mathrm}
  \DeclareOldFontCommand{\sf}{\normalfont\sffamily}{\mathsf}
  \DeclareOldFontCommand{\tt}{\normalfont\ttfamily}{\mathtt}
  \DeclareOldFontCommand{\bf}{\normalfont\bfseries}{\mathbf}
  \DeclareOldFontCommand{\it}{\normalfont\itshape}{\mathit}
  \DeclareOldFontCommand{\sl}{\normalfont\slshape}{\@nomath\sl}
  \DeclareOldFontCommand{\sc}{\normalfont\scshape}{\@nomath\sc}
  \chkcompat
\fi

\def\alpha{\Greekmath 010B }%
\def\beta{\Greekmath 010C }%
\def\gamma{\Greekmath 010D }%
\def\delta{\Greekmath 010E }%
\def\epsilon{\Greekmath 010F }%
\def\zeta{\Greekmath 0110 }%
\def\eta{\Greekmath 0111 }%
\def\theta{\Greekmath 0112 }%
\def\iota{\Greekmath 0113 }%
\def\kappa{\Greekmath 0114 }%
\def\lambda{\Greekmath 0115 }%
\def\mu{\Greekmath 0116 }%
\def\nu{\Greekmath 0117 }%
\def\xi{\Greekmath 0118 }%
\def\pi{\Greekmath 0119 }%
\def\rho{\Greekmath 011A }%
\def\sigma{\Greekmath 011B }%
\def\tau{\Greekmath 011C }%
\def\upsilon{\Greekmath 011D }%
\def\phi{\Greekmath 011E }%
\def\chi{\Greekmath 011F }%
\def\psi{\Greekmath 0120 }%
\def\omega{\Greekmath 0121 }%
\def\varepsilon{\Greekmath 0122 }%
\def\vartheta{\Greekmath 0123 }%
\def\varpi{\Greekmath 0124 }%
\def\varrho{\Greekmath 0125 }%
\def\varsigma{\Greekmath 0126 }%
\def\varphi{\Greekmath 0127 }%

\def\nabla{\Greekmath 0272 }

\def\Greekmath#1#2#3#4{%
    \if@compatibility
        \ifnum\mathgroup=\symbold
           \mathchoice{\mbox{\boldmath$\displaystyle\mathchar"#1#2#3#4$}}%
                      {\mbox{\boldmath$\textstyle\mathchar"#1#2#3#4$}}%
                      {\mbox{\boldmath$\scriptstyle\mathchar"#1#2#3#4$}}%
                      {\mbox{\boldmath$\scriptscriptstyle\mathchar"#1#2#3#4$}}%
        \else
           \mathchar"#1#2#3#4%
        \fi 
    \else 
        \ifnum\mathgroup=5 
           \mathchoice{\mbox{\boldmath$\displaystyle\mathchar"#1#2#3#4$}}%
                      {\mbox{\boldmath$\textstyle\mathchar"#1#2#3#4$}}%
                      {\mbox{\boldmath$\scriptstyle\mathchar"#1#2#3#4$}}%
                      {\mbox{\boldmath$\scriptscriptstyle\mathchar"#1#2#3#4$}}%
        \else
           \mathchar"#1#2#3#4%
        \fi     	    
	  \fi}

\newif\ifGreekBold  \GreekBoldfalse
\let\SAVEPBF=\pbf
\def\pbf{\GreekBoldtrue\SAVEPBF}%

\@ifundefined{theorem}{\newtheorem{theorem}{Theorem}}{}
\@ifundefined{lemma}{\newtheorem{lemma}[theorem]{Lemma}}{}
\@ifundefined{corollary}{}{}
\@ifundefined{conjecture}{}{}
\@ifundefined{proposition}{\newtheorem{proposition}[theorem]{Proposition}}{}
\@ifundefined{axiom}{}{}
\@ifundefined{remark}{}{}
\@ifundefined{example}{}{}
\@ifundefined{exercise}{}{}
\@ifundefined{definition}{\newtheorem{definition}{Definition}}{}

\@ifundefined{mathletters}{%
  \newcounter{equationnumber}  
  \def\mathletters{%
     \addtocounter{equation}{1}
     \edef\@currentlabel{\theequation}%
     \setcounter{equationnumber}{\c@equation}
     \setcounter{equation}{0}%
     \edef\theequation{\@currentlabel\noexpand\alph{equation}}%
  }
  
}{}

\@ifundefined{BibTeX}{%
    \def\BibTeX{{\rm B\kern-.05em{\sc i\kern-.025em b}\kern-.08em
                 T\kern-.1667em\lower.7ex\hbox{E}\kern-.125emX}}}{}%
\@ifundefined{AmS}%
    {\def\AmS{{\protect\usefont{OMS}{cmsy}{m}{n}%
                A\kern-.1667em\lower.5ex\hbox{M}\kern-.125emS}}}{}%
\@ifundefined{AmSTeX}{}{}%
\ifx\ds@amstex\relax
\makeatother\endinput
\else
   \@ifpackageloaded{amstex}%
      {\message{amstex already loaded}\makeatother\endinput}
      {}
\fi
\let\DOTSI\relax
\def\RIfM@{\relax\ifmmode}%
\def\FN@{\futurelet\next}%
\newcount\intno@
\def\iint{\DOTSI\intno@\tw@\FN@\ints@}%
\def\iiint{\DOTSI\intno@\thr@@\FN@\ints@}%
\def\iiiint{\DOTSI\intno@4 \FN@\ints@}%
\def\idotsint{\DOTSI\intno@\z@\FN@\ints@}%
\def\ints@{\findlimits@\ints@@}%
\newif\iflimtoken@
\newif\iflimits@
\def\findlimits@{\limtoken@true\ifx\next\limits\limits@true
 \else\ifx\next\nolimits\limits@false\else
 \limtoken@false\ifx\ilimits@\nolimits\limits@false\else
 \ifinner\limits@false\else\limits@true\fi\fi\fi\fi}%
\def\multint@{\int\ifnum\intno@=\z@\intdots@                          
 \else\intkern@\fi                                                    
 \ifnum\intno@>\tw@\int\intkern@\fi                                   
 \ifnum\intno@>\thr@@\int\intkern@\fi                                 
 \int}
\def\multintlimits@{\intop\ifnum\intno@=\z@\intdots@\else\intkern@\fi
 \ifnum\intno@>\tw@\intop\intkern@\fi
 \ifnum\intno@>\thr@@\intop\intkern@\fi\intop}%
\def\intic@{%
    \mathchoice{\hskip.5em}{\hskip.4em}{\hskip.4em}{\hskip.4em}}%
\def\negintic@{\mathchoice
 {\hskip-.5em}{\hskip-.4em}{\hskip-.4em}{\hskip-.4em}}%
\def\ints@@{\iflimtoken@                                              
 \def\ints@@@{\iflimits@\negintic@
   \mathop{\intic@\multintlimits@}\limits                             
  \else\multint@\nolimits\fi                                          
  \eat@}
 \else                                                                
 \def\ints@@@{\iflimits@\negintic@
  \mathop{\intic@\multintlimits@}\limits\else
  \multint@\nolimits\fi}\fi\ints@@@}%
\def\intkern@{\mathchoice{\!\!\!}{\!\!}{\!\!}{\!\!}}%
\def\plaincdots@{\mathinner{\cdotp\cdotp\cdotp}}%
\def\intdots@{\mathchoice{\plaincdots@}%
 {{\cdotp}\mkern1.5mu{\cdotp}\mkern1.5mu{\cdotp}}%
 {{\cdotp}\mkern1mu{\cdotp}\mkern1mu{\cdotp}}%
 {{\cdotp}\mkern1mu{\cdotp}\mkern1mu{\cdotp}}}%
\def\RIfM@{\relax\protect\ifmmode}
\def\text{\RIfM@\expandafter\text@\else\expandafter\mbox\fi}
\let\nfss@text\text
\def\text@#1{\mathchoice
   {\textdef@\displaystyle\f@size{#1}}%
   {\textdef@\textstyle\tf@size{\firstchoice@false #1}}%
   {\textdef@\textstyle\sf@size{\firstchoice@false #1}}%
   {\textdef@\textstyle \ssf@size{\firstchoice@false #1}}%
   \glb@settings}

\def\textdef@#1#2#3{\hbox{{%
                    \everymath{#1}%
                    \let\f@size#2\selectfont
                    #3}}}
\newif\iffirstchoice@
\firstchoice@true
\def\Let@{\relax\iffalse{\fi\let\\=\cr\iffalse}\fi}%
\def\vspace@{\def\vspace##1{\crcr\noalign{\vskip##1\relax}}}%
\def\multilimits@{\bgroup\vspace@\Let@
 \baselineskip\fontdimen10 \scriptfont\tw@
 \advance\baselineskip\fontdimen12 \scriptfont\tw@
 \lineskip\thr@@\fontdimen8 \scriptfont\thr@@
 \lineskiplimit\lineskip
 \vbox\bgroup\ialign\bgroup\hfil$\m@th\scriptstyle{##}$\hfil\crcr}%
\def\Sb{_\multilimits@}%
\def\endSb{\crcr\egroup\egroup\egroup}%
\def\Sp{^\multilimits@}%

\newdimen\ex@
\def\rightarrowfill@#1{$#1\m@th\mathord-\mkern-6mu\cleaders
 \hbox{$#1\mkern-2mu\mathord-\mkern-2mu$}\hfill
 \mkern-6mu\mathord\rightarrow$}%
\def\leftarrowfill@#1{$#1\m@th\mathord\leftarrow\mkern-6mu\cleaders
 \hbox{$#1\mkern-2mu\mathord-\mkern-2mu$}\hfill\mkern-6mu\mathord-$}%
\def\leftrightarrowfill@#1{$#1\m@th\mathord\leftarrow
\mkern-6mu\cleaders
 \hbox{$#1\mkern-2mu\mathord-\mkern-2mu$}\hfill
 \mkern-6mu\mathord\rightarrow$}%
\def\overrightarrow{\mathpalette\overrightarrow@}%
\def\overrightarrow@#1#2{\vbox{\ialign{##\crcr\rightarrowfill@#1\crcr
 \noalign{\kern-\ex@\nointerlineskip}$\m@th\hfil#1#2\hfil$\crcr}}}%

\def\overleftarrow{\mathpalette\overleftarrow@}%
\def\overleftarrow@#1#2{\vbox{\ialign{##\crcr\leftarrowfill@#1\crcr
 \noalign{\kern-\ex@\nointerlineskip}$\m@th\hfil#1#2\hfil$\crcr}}}%
\def\overleftrightarrow{\mathpalette\overleftrightarrow@}%
\def\overleftrightarrow@#1#2{\vbox{\ialign{##\crcr
   \leftrightarrowfill@#1\crcr
 \noalign{\kern-\ex@\nointerlineskip}$\m@th\hfil#1#2\hfil$\crcr}}}%
\def\underrightarrow{\mathpalette\underrightarrow@}%
\def\underrightarrow@#1#2{\vtop{\ialign{##\crcr$\m@th\hfil#1#2\hfil
  $\crcr\noalign{\nointerlineskip}\rightarrowfill@#1\crcr}}}%

\def\underleftarrow{\mathpalette\underleftarrow@}%
\def\underleftarrow@#1#2{\vtop{\ialign{##\crcr$\m@th\hfil#1#2\hfil
  $\crcr\noalign{\nointerlineskip}\leftarrowfill@#1\crcr}}}%
\def\underleftrightarrow{\mathpalette\underleftrightarrow@}%
\def\underleftrightarrow@#1#2{\vtop{\ialign{##\crcr$\m@th
  \hfil#1#2\hfil$\crcr
 \noalign{\nointerlineskip}\leftrightarrowfill@#1\crcr}}}%

\def\qopnamewl@#1{\mathop{\operator@font#1}\nlimits@}
\let\nlimits@\displaylimits
\def\setboxz@h{\setbox\z@\hbox}

\def\varlim@#1#2{\mathop{\vtop{\ialign{##\crcr
 \hfil$#1\m@th\operator@font lim$\hfil\crcr
 \noalign{\nointerlineskip}#2#1\crcr
 \noalign{\nointerlineskip\kern-\ex@}\crcr}}}}

 \def\rightarrowfill@#1{\m@th\setboxz@h{$#1-$}\ht\z@\z@
  $#1\copy\z@\mkern-6mu\cleaders
  \hbox{$#1\mkern-2mu\box\z@\mkern-2mu$}\hfill
  \mkern-6mu\mathord\rightarrow$}
\def\leftarrowfill@#1{\m@th\setboxz@h{$#1-$}\ht\z@\z@
  $#1\mathord\leftarrow\mkern-6mu\cleaders
  \hbox{$#1\mkern-2mu\copy\z@\mkern-2mu$}\hfill
  \mkern-6mu\box\z@$}

\def\projlim{\qopnamewl@{proj\,lim}}
\def\injlim{\qopnamewl@{inj\,lim}}
\def\varinjlim{\mathpalette\varlim@\rightarrowfill@}
\def\varprojlim{\mathpalette\varlim@\leftarrowfill@}
\def\varliminf{\mathpalette\varliminf@{}}
\def\varliminf@#1{\mathop{\underline{\vrule\@depth.2\ex@\@width\z@
   \hbox{$#1\m@th\operator@font lim$}}}}
\def\varlimsup{\mathpalette\varlimsup@{}}
\def\varlimsup@#1{\mathop{\overline
  {\hbox{$#1\m@th\operator@font lim$}}}}

\begingroup \catcode `|=0 \catcode `[= 1
\catcode`]=2 \catcode `\{=12 \catcode `\}=12
\catcode`\\=12 
|gdef|@alignverbatim#1\end{align}[#1|end[align]]
|gdef|@salignverbatim#1\end{align*}[#1|end[align*]]

|gdef|@alignatverbatim#1\end{alignat}[#1|end[alignat]]
|gdef|@salignatverbatim#1\end{alignat*}[#1|end[alignat*]]

|gdef|@xalignatverbatim#1\end{xalignat}[#1|end[xalignat]]
|gdef|@sxalignatverbatim#1\end{xalignat*}[#1|end[xalignat*]]

|gdef|@gatherverbatim#1\end{gather}[#1|end[gather]]
|gdef|@sgatherverbatim#1\end{gather*}[#1|end[gather*]]

|gdef|@gatherverbatim#1\end{gather}[#1|end[gather]]
|gdef|@sgatherverbatim#1\end{gather*}[#1|end[gather*]]

|gdef|@multilineverbatim#1\end{multiline}[#1|end[multiline]]
|gdef|@smultilineverbatim#1\end{multiline*}[#1|end[multiline*]]

|gdef|@arraxverbatim#1\end{arrax}[#1|end[arrax]]
|gdef|@sarraxverbatim#1\end{arrax*}[#1|end[arrax*]]

|gdef|@tabulaxverbatim#1\end{tabulax}[#1|end[tabulax]]
|gdef|@stabulaxverbatim#1\end{tabulax*}[#1|end[tabulax*]]

|endgroup

\def\align{\@verbatim \frenchspacing\@vobeyspaces \@alignverbatim
You are using the "align" environment in a style in which it is not defined.}

\@namedef{align*}{\@verbatim\@salignverbatim
You are using the "align*" environment in a style in which it is not defined.}
\expandafter\let\csname endalign*\endcsname =\endtrivlist

\def\alignat{\@verbatim \frenchspacing\@vobeyspaces \@alignatverbatim
You are using the "alignat" environment in a style in which it is not defined.}

\@namedef{alignat*}{\@verbatim\@salignatverbatim
You are using the "alignat*" environment in a style in which it is not defined.}
\expandafter\let\csname endalignat*\endcsname =\endtrivlist

\def\xalignat{\@verbatim \frenchspacing\@vobeyspaces \@xalignatverbatim
You are using the "xalignat" environment in a style in which it is not defined.}

\@namedef{xalignat*}{\@verbatim\@sxalignatverbatim
You are using the "xalignat*" environment in a style in which it is not defined.}
\expandafter\let\csname endxalignat*\endcsname =\endtrivlist

\def\gather{\@verbatim \frenchspacing\@vobeyspaces \@gatherverbatim
You are using the "gather" environment in a style in which it is not defined.}

\@namedef{gather*}{\@verbatim\@sgatherverbatim
You are using the "gather*" environment in a style in which it is not defined.}
\expandafter\let\csname endgather*\endcsname =\endtrivlist

\def\multiline{\@verbatim \frenchspacing\@vobeyspaces \@multilineverbatim
You are using the "multiline" environment in a style in which it is not defined.}

\@namedef{multiline*}{\@verbatim\@smultilineverbatim
You are using the "multiline*" environment in a style in which it is not defined.}
\expandafter\let\csname endmultiline*\endcsname =\endtrivlist

\def\arrax{\@verbatim \frenchspacing\@vobeyspaces \@arraxverbatim
You are using a type of "array" construct that is only allowed in AmS-LaTeX.}

\def\tabulax{\@verbatim \frenchspacing\@vobeyspaces \@tabulaxverbatim
You are using a type of "tabular" construct that is only allowed in AmS-LaTeX.}

\@namedef{arrax*}{\@verbatim\@sarraxverbatim
You are using a type of "array*" construct that is only allowed in AmS-LaTeX.}
\expandafter\let\csname endarrax*\endcsname =\endtrivlist

\@namedef{tabulax*}{\@verbatim\@stabulaxverbatim
You are using a type of "tabular*" construct that is only allowed in AmS-LaTeX.}
\expandafter\let\csname endtabulax*\endcsname =\endtrivlist

\def\@@eqncr{\let\@tempa\relax
    \ifcase\@eqcnt \def\@tempa{& & &}\or \def\@tempa{& &}%
      \else \def\@tempa{&}\fi
     \@tempa
     \if@eqnsw
        \iftag@
           \@taggnum
        \else
           \@eqnnum\stepcounter{equation}%
        \fi
     \fi
     \global\tag@false
     \global\@eqnswtrue
     \global\@eqcnt\z@\cr}

 \def\endequation{%
     \ifmmode\ifinner 
      \iftag@
        \addtocounter{equation}{-1} 
        $\hfil
           \displaywidth\linewidth\@taggnum\egroup \endtrivlist
        \global\tag@false
        \global\@ignoretrue   
      \else
        $\hfil
           \displaywidth\linewidth\@eqnnum\egroup \endtrivlist
        \global\tag@false
        \global\@ignoretrue 
      \fi
     \else   
      \iftag@
        \addtocounter{equation}{-1} 
        \eqno \hbox{\@taggnum}
        \global\tag@false%
        $$\global\@ignoretrue
      \else
        \eqno \hbox{\@eqnnum}
        $$\global\@ignoretrue
      \fi
     \fi\fi
 } 

 \newif\iftag@ \tag@false
 
 \def\tag{\@ifnextchar*{\@tagstar}{\@tag}}
 \def\@tag#1{%
     \global\tag@true
     \global\def\@taggnum{(#1)}}
 \def\@tagstar*#1{%
     \global\tag@true
     \global\def\@taggnum{#1}%
}

\makeatother

\catcode `@=11

\let\@amsfonts=T
\RequirePackage{amsgen}
\newbox\Mathstrutbox@
\setbox\Mathstrutbox@=\hbox{}
\def\Mathstrut@{\copy\Mathstrutbox@}
\addto@hook\every@math@size{\setbox\z@\hbox{\normalfont(}%
  \ht\Mathstrutbox@\ht\z@ \dp\Mathstrutbox@\dp\z@}
\newbox\strutbox@
\def\strut@{\copy\strutbox@}
\addto@hook\every@math@size{%
  \global\setbox\strutbox@\hbox{\lower.5\normallineskiplimit
         \vbox{\kern-\normallineskiplimit\copy\strutbox}}}
\def\big{\bBigg@\@ne}
\def\Big{\bBigg@{1.5}}
\def\bigg{\bBigg@\tw@}
\def\Bigg{\bBigg@{2.5}}
\def\bBigg@#1#2{%
   {%
    \hbox{$\left#2\vcenter to#1\big@size{}\right.%
           \n@space
     $}}}
\addto@hook\every@math@size{%
  \global\big@size 1.2\ht\Mathstrutbox@
  \global\advance\big@size 1.2\dp\Mathstrutbox@ }
\newdimen\big@size
\ifx F\@amsfonts \endinput\fi
\DeclareSymbolFont{AMSa}{U}{msa}{m}{n}
\DeclareSymbolFont{AMSb}{U}{msb}{m}{n}
\@ifundefined{yen}{%
  \edef\yen{\noexpand\mathhexbox{\hexnumber@\symAMSa}55}
}{}
\@ifundefined{checkmark}{%
  \edef\checkmark{\noexpand\mathhexbox{\hexnumber@\symAMSa}58}
}{}
\@ifundefined{circledR}{%
  \edef\circledR{\noexpand\mathhexbox{\hexnumber@\symAMSa}72}
}{}
\@ifundefined{maltese}{%
  \edef\maltese{\noexpand\mathhexbox{\hexnumber@\symAMSa}7A}
}{}
\begingroup \catcode`\"=12
\DeclareMathDelimiter\ulcorner{\mathopen} {AMSa}{"70}{AMSa}{"70}
\DeclareMathDelimiter\urcorner{\mathclose}{AMSa}{"71}{AMSa}{"71}
\DeclareMathDelimiter\llcorner{\mathopen} {AMSa}{"78}{AMSa}{"78}
\DeclareMathDelimiter\lrcorner{\mathclose}{AMSa}{"79}{AMSa}{"79}
\xdef\widehat#1{\noexpand\@mathmeasure\z@\textstyle{#1}%
  \noexpand\ifdim\noexpand\wdz@>\tw@ em%
  \mathaccent"0\hexnumber@\symAMSb 5B{#1}%
  \noexpand\else\mathaccent"0362{#1}\noexpand\fi}
\xdef\widetilde#1{\noexpand\@mathmeasure\z@\textstyle{#1}%
  \noexpand\ifdim\noexpand\wdz@>\tw@ em%
  \mathaccent"0\hexnumber@\symAMSb 5D{#1}%
  \noexpand\else\mathaccent"0365{#1}\noexpand\fi}
\DeclareMathSymbol\dabar@{\mathord}{AMSa}{"39}
\xdef\dashrightarrow{\mathrel{\dabar@\dabar@
                              \mathchar"0\hexnumber@\symAMSa 4B}}%
\xdef\dashleftarrow{\mathrel{\mathchar"0\hexnumber@\symAMSa 4C\dabar@
                              \dabar@}}%

\global\let\rightleftharpoons\undefined
\DeclareMathSymbol\rightleftharpoons{\mathrel}{AMSa}{"0A}
\global\let\angle\undefined
\DeclareMathSymbol\angle            {\mathord}{AMSa}{"5C}
\global\let\hbar\undefined
\DeclareMathSymbol\hbar             {\mathord}{AMSb}{"7E}
\global\let\sqsubset\undefined
\DeclareMathSymbol\sqsubset         {\mathrel}{AMSa}{"40}
\global\let\sqsupset\undefined
\DeclareMathSymbol\sqsupset         {\mathrel}{AMSa}{"41}
\global\let\mho\undefined
\DeclareMathSymbol\mho              {\mathord}{AMSb}{"66}
\DeclareMathSymbol\square           {\mathord}{AMSa}{"03}
\DeclareMathSymbol\lozenge          {\mathord}{AMSa}{"06}
\DeclareMathSymbol\vartriangleright {\mathrel}{AMSa}{"42}
\DeclareMathSymbol\vartriangleleft  {\mathrel}{AMSa}{"43}
\DeclareMathSymbol\trianglerighteq  {\mathrel}{AMSa}{"44}
\DeclareMathSymbol\trianglelefteq   {\mathrel}{AMSa}{"45}
\DeclareMathSymbol\rightsquigarrow  {\mathrel}{AMSa}{"20}
\def\@tempa{\not@base\lhd}
\ifx\lhd\@tempa
 \global\let\lhd\vartriangleleft

 \global\let\leadsto\rightsquigarrow
\xdef\Join{\mathrel{\mathchar"0\hexnumber@\symAMSb 6F\mkern-13.8mu%
  \mathchar"0\hexnumber@\symAMSb 6E}}
\fi
\endgroup
\DeclareMathAlphabet\mathfrak{U}{euf}{m}{n}
\SetMathAlphabet\mathfrak{bold}{U}{euf}{b}{n}
\DeclareSymbolFontAlphabet{\mathbb}{AMSb}
\DeclareFontEncodingDefaults{\relax}{\def\accentclass@{7}}
\def\frak{\@subst@obsolete\frak\mathfrak}
\def\Bbb{\@subst@obsolete\Bbb\mathbb}
\def\bold{\@subst@obsolete\bold\mathbf}
\begingroup \catcode`\"=12 \relax
\gdef\newsymbol#1#2#3#4#5{%
  \@obsolete\newsymbol\DeclareMathSymbol
  \@ifdefinable#1{%
     \edef\next@
       {\ifcase #2 \or
          \hexnumber@\symAMSa\or
          \hexnumber@\symAMSb\fi}%
     \ifx\next@\@empty
       \PackageError{amsfonts}{\Invalid@@\newsymbol}\@ehd%
     \else
      \global\mathchardef#1"#3\next@#4#5
     \fi}}
\endgroup
\RequirePackage{amsfonts}
\DeclareMathSymbol{\boxdot}       {\mathbin}{AMSa}{"00}
\DeclareMathSymbol{\boxplus}      {\mathbin}{AMSa}{"01}
\DeclareMathSymbol{\boxtimes}     {\mathbin}{AMSa}{"02}
\DeclareMathSymbol{\square}       {\mathord}{AMSa}{"03}
\DeclareMathSymbol{\blacksquare}  {\mathord}{AMSa}{"04}
\DeclareMathSymbol{\centerdot}    {\mathbin}{AMSa}{"05}
\DeclareMathSymbol{\lozenge}      {\mathord}{AMSa}{"06}
\DeclareMathSymbol{\blacklozenge} {\mathord}{AMSa}{"07}
\DeclareMathSymbol{\circlearrowright}   {\mathrel}{AMSa}{"08}
\DeclareMathSymbol{\circlearrowleft}    {\mathrel}{AMSa}{"09}
\DeclareMathSymbol{\leftrightharpoons}  {\mathrel}{AMSa}{"0B}
\DeclareMathSymbol{\boxminus}     {\mathbin}{AMSa}{"0C}
\DeclareMathSymbol{\Vdash}        {\mathrel}{AMSa}{"0D}
\DeclareMathSymbol{\Vvdash}       {\mathrel}{AMSa}{"0E}
\DeclareMathSymbol{\vDash}        {\mathrel}{AMSa}{"0F}
\DeclareMathSymbol{\twoheadrightarrow}  {\mathrel}{AMSa}{"10}
\DeclareMathSymbol{\twoheadleftarrow}   {\mathrel}{AMSa}{"11}
\DeclareMathSymbol{\leftleftarrows}     {\mathrel}{AMSa}{"12}
\DeclareMathSymbol{\rightrightarrows}   {\mathrel}{AMSa}{"13}
\DeclareMathSymbol{\upuparrows}         {\mathrel}{AMSa}{"14}
\DeclareMathSymbol{\downdownarrows} {\mathrel}{AMSa}{"15}
\DeclareMathSymbol{\upharpoonright} {\mathrel}{AMSa}{"16}
 
\DeclareMathSymbol{\downharpoonright}   {\mathrel}{AMSa}{"17}
\DeclareMathSymbol{\upharpoonleft}  {\mathrel}{AMSa}{"18}
\DeclareMathSymbol{\downharpoonleft}{\mathrel}{AMSa}{"19}
\DeclareMathSymbol{\rightarrowtail} {\mathrel}{AMSa}{"1A}
\DeclareMathSymbol{\leftarrowtail}  {\mathrel}{AMSa}{"1B}
\DeclareMathSymbol{\leftrightarrows}{\mathrel}{AMSa}{"1C}
\DeclareMathSymbol{\rightleftarrows}{\mathrel}{AMSa}{"1D}
\DeclareMathSymbol{\Lsh}            {\mathrel}{AMSa}{"1E}
\DeclareMathSymbol{\Rsh}            {\mathrel}{AMSa}{"1F}
\DeclareMathSymbol{\leftrightsquigarrow}{\mathrel}{AMSa}{"21}
\DeclareMathSymbol{\looparrowleft}  {\mathrel}{AMSa}{"22}
\DeclareMathSymbol{\looparrowright} {\mathrel}{AMSa}{"23}
\DeclareMathSymbol{\circeq}       {\mathrel}{AMSa}{"24}
\DeclareMathSymbol{\succsim}      {\mathrel}{AMSa}{"25}
\DeclareMathSymbol{\gtrsim}       {\mathrel}{AMSa}{"26}
\DeclareMathSymbol{\gtrapprox}    {\mathrel}{AMSa}{"27}
\DeclareMathSymbol{\multimap}     {\mathrel}{AMSa}{"28}
\DeclareMathSymbol{\because}      {\mathrel}{AMSa}{"2A}
\DeclareMathSymbol{\doteqdot}     {\mathrel}{AMSa}{"2B}
 
\DeclareMathSymbol{\triangleq}    {\mathrel}{AMSa}{"2C}
\DeclareMathSymbol{\precsim}      {\mathrel}{AMSa}{"2D}
\DeclareMathSymbol{\lesssim}      {\mathrel}{AMSa}{"2E}
\DeclareMathSymbol{\lessapprox}   {\mathrel}{AMSa}{"2F}
\DeclareMathSymbol{\eqslantless}  {\mathrel}{AMSa}{"30}
\DeclareMathSymbol{\eqslantgtr}   {\mathrel}{AMSa}{"31}
\DeclareMathSymbol{\curlyeqprec}  {\mathrel}{AMSa}{"32}
\DeclareMathSymbol{\curlyeqsucc}  {\mathrel}{AMSa}{"33}
\DeclareMathSymbol{\preccurlyeq}  {\mathrel}{AMSa}{"34}
\DeclareMathSymbol{\leqq}         {\mathrel}{AMSa}{"35}
\DeclareMathSymbol{\leqslant}     {\mathrel}{AMSa}{"36}
\DeclareMathSymbol{\lessgtr}      {\mathrel}{AMSa}{"37}
\DeclareMathSymbol{\backprime}    {\mathord}{AMSa}{"38}
\DeclareMathSymbol{\risingdotseq} {\mathrel}{AMSa}{"3A}
\DeclareMathSymbol{\fallingdotseq}{\mathrel}{AMSa}{"3B}
\DeclareMathSymbol{\succcurlyeq}  {\mathrel}{AMSa}{"3C}
\DeclareMathSymbol{\geqq}         {\mathrel}{AMSa}{"3D}
\DeclareMathSymbol{\geqslant}     {\mathrel}{AMSa}{"3E}
\DeclareMathSymbol{\gtrless}      {\mathrel}{AMSa}{"3F}
\DeclareMathSymbol{\bigstar}    {\mathord}{AMSa}{"46}
\DeclareMathSymbol{\between}    {\mathrel}{AMSa}{"47}
\DeclareMathSymbol{\blacktriangledown}  {\mathord}{AMSa}{"48}
\DeclareMathSymbol{\blacktriangleright} {\mathrel}{AMSa}{"49}
\DeclareMathSymbol{\blacktriangleleft}  {\mathrel}{AMSa}{"4A}
\DeclareMathSymbol{\vartriangle}        {\mathrel}{AMSa}{"4D}
\DeclareMathSymbol{\blacktriangle}      {\mathord}{AMSa}{"4E}
\DeclareMathSymbol{\triangledown}       {\mathord}{AMSa}{"4F}
\DeclareMathSymbol{\eqcirc}       {\mathrel}{AMSa}{"50}
\DeclareMathSymbol{\lesseqgtr}    {\mathrel}{AMSa}{"51}
\DeclareMathSymbol{\gtreqless}    {\mathrel}{AMSa}{"52}
\DeclareMathSymbol{\lesseqqgtr}   {\mathrel}{AMSa}{"53}
\DeclareMathSymbol{\gtreqqless}   {\mathrel}{AMSa}{"54}
\DeclareMathSymbol{\Rrightarrow}  {\mathrel}{AMSa}{"56}
\DeclareMathSymbol{\Lleftarrow}   {\mathrel}{AMSa}{"57}
\DeclareMathSymbol{\veebar}       {\mathbin}{AMSa}{"59}
\DeclareMathSymbol{\barwedge}     {\mathbin}{AMSa}{"5A}
\DeclareMathSymbol{\doublebarwedge} {\mathbin}{AMSa}{"5B}
\DeclareMathSymbol{\measuredangle}  {\mathord}{AMSa}{"5D}
\DeclareMathSymbol{\sphericalangle} {\mathord}{AMSa}{"5E}
\DeclareMathSymbol{\varpropto}    {\mathrel}{AMSa}{"5F}
\DeclareMathSymbol{\smallsmile}   {\mathrel}{AMSa}{"60}
\DeclareMathSymbol{\smallfrown}   {\mathrel}{AMSa}{"61}
\DeclareMathSymbol{\Subset}       {\mathrel}{AMSa}{"62}
\DeclareMathSymbol{\Supset}       {\mathrel}{AMSa}{"63}
\DeclareMathSymbol{\Cup}          {\mathbin}{AMSa}{"64}
 
\DeclareMathSymbol{\Cap}          {\mathbin}{AMSa}{"65}
 
\DeclareMathSymbol{\curlywedge}   {\mathbin}{AMSa}{"66}
\DeclareMathSymbol{\curlyvee}     {\mathbin}{AMSa}{"67}
\DeclareMathSymbol{\leftthreetimes} {\mathbin}{AMSa}{"68}
\DeclareMathSymbol{\rightthreetimes}{\mathbin}{AMSa}{"69}
\DeclareMathSymbol{\subseteqq}    {\mathrel}{AMSa}{"6A}
\DeclareMathSymbol{\supseteqq}    {\mathrel}{AMSa}{"6B}
\DeclareMathSymbol{\bumpeq}       {\mathrel}{AMSa}{"6C}
\DeclareMathSymbol{\Bumpeq}       {\mathrel}{AMSa}{"6D}
\DeclareMathSymbol{\lll}          {\mathrel}{AMSa}{"6E}
 
\DeclareMathSymbol{\ggg}          {\mathrel}{AMSa}{"6F}
 
\DeclareMathSymbol{\circledS}     {\mathord}{AMSa}{"73}
\DeclareMathSymbol{\pitchfork}    {\mathrel}{AMSa}{"74}
\DeclareMathSymbol{\dotplus}      {\mathbin}{AMSa}{"75}
\DeclareMathSymbol{\backsim}      {\mathrel}{AMSa}{"76}
\DeclareMathSymbol{\backsimeq}    {\mathrel}{AMSa}{"77}
\DeclareMathSymbol{\complement}   {\mathord}{AMSa}{"7B}
\DeclareMathSymbol{\intercal}     {\mathbin}{AMSa}{"7C}
\DeclareMathSymbol{\circledcirc}  {\mathbin}{AMSa}{"7D}
\DeclareMathSymbol{\circledast}   {\mathbin}{AMSa}{"7E}
\DeclareMathSymbol{\circleddash}  {\mathbin}{AMSa}{"7F}
\DeclareMathSymbol{\lvertneqq}    {\mathrel}{AMSb}{"00}
\DeclareMathSymbol{\gvertneqq}    {\mathrel}{AMSb}{"01}
\DeclareMathSymbol{\nleq}         {\mathrel}{AMSb}{"02}
\DeclareMathSymbol{\ngeq}         {\mathrel}{AMSb}{"03}
\DeclareMathSymbol{\nless}        {\mathrel}{AMSb}{"04}
\DeclareMathSymbol{\ngtr}         {\mathrel}{AMSb}{"05}
\DeclareMathSymbol{\nprec}        {\mathrel}{AMSb}{"06}
\DeclareMathSymbol{\nsucc}        {\mathrel}{AMSb}{"07}
\DeclareMathSymbol{\lneqq}        {\mathrel}{AMSb}{"08}
\DeclareMathSymbol{\gneqq}        {\mathrel}{AMSb}{"09}
\DeclareMathSymbol{\nleqslant}    {\mathrel}{AMSb}{"0A}
\DeclareMathSymbol{\ngeqslant}    {\mathrel}{AMSb}{"0B}
\DeclareMathSymbol{\lneq}         {\mathrel}{AMSb}{"0C}
\DeclareMathSymbol{\gneq}         {\mathrel}{AMSb}{"0D}
\DeclareMathSymbol{\npreceq}      {\mathrel}{AMSb}{"0E}
\DeclareMathSymbol{\nsucceq}      {\mathrel}{AMSb}{"0F}
\DeclareMathSymbol{\precnsim}     {\mathrel}{AMSb}{"10}
\DeclareMathSymbol{\succnsim}     {\mathrel}{AMSb}{"11}
\DeclareMathSymbol{\lnsim}        {\mathrel}{AMSb}{"12}
\DeclareMathSymbol{\gnsim}        {\mathrel}{AMSb}{"13}
\DeclareMathSymbol{\nleqq}        {\mathrel}{AMSb}{"14}
\DeclareMathSymbol{\ngeqq}        {\mathrel}{AMSb}{"15}
\DeclareMathSymbol{\precneqq}     {\mathrel}{AMSb}{"16}
\DeclareMathSymbol{\succneqq}     {\mathrel}{AMSb}{"17}
\DeclareMathSymbol{\precnapprox}  {\mathrel}{AMSb}{"18}
\DeclareMathSymbol{\succnapprox}  {\mathrel}{AMSb}{"19}
\DeclareMathSymbol{\lnapprox}     {\mathrel}{AMSb}{"1A}
\DeclareMathSymbol{\gnapprox}     {\mathrel}{AMSb}{"1B}
\DeclareMathSymbol{\nsim}         {\mathrel}{AMSb}{"1C}
\DeclareMathSymbol{\ncong}        {\mathrel}{AMSb}{"1D}
\DeclareMathSymbol{\diagup}       {\mathrel}{AMSb}{"1E}
\DeclareMathSymbol{\diagdown}     {\mathrel}{AMSb}{"1F}
\DeclareMathSymbol{\varsubsetneq}   {\mathrel}{AMSb}{"20}
\DeclareMathSymbol{\varsupsetneq}   {\mathrel}{AMSb}{"21}
\DeclareMathSymbol{\nsubseteqq}     {\mathrel}{AMSb}{"22}
\DeclareMathSymbol{\nsupseteqq}     {\mathrel}{AMSb}{"23}
\DeclareMathSymbol{\subsetneqq}     {\mathrel}{AMSb}{"24}
\DeclareMathSymbol{\supsetneqq}     {\mathrel}{AMSb}{"25}
\DeclareMathSymbol{\varsubsetneqq}  {\mathrel}{AMSb}{"26}
\DeclareMathSymbol{\varsupsetneqq}  {\mathrel}{AMSb}{"27}
\DeclareMathSymbol{\subsetneq}      {\mathrel}{AMSb}{"28}
\DeclareMathSymbol{\supsetneq}      {\mathrel}{AMSb}{"29}
\DeclareMathSymbol{\nsubseteq}      {\mathrel}{AMSb}{"2A}
\DeclareMathSymbol{\nsupseteq}      {\mathrel}{AMSb}{"2B}
\DeclareMathSymbol{\nparallel}      {\mathrel}{AMSb}{"2C}
\DeclareMathSymbol{\nmid}           {\mathrel}{AMSb}{"2D}
\DeclareMathSymbol{\nshortmid}      {\mathrel}{AMSb}{"2E}
\DeclareMathSymbol{\nshortparallel} {\mathrel}{AMSb}{"2F}
\DeclareMathSymbol{\nvdash}         {\mathrel}{AMSb}{"30}
\DeclareMathSymbol{\nVdash}         {\mathrel}{AMSb}{"31}
\DeclareMathSymbol{\nvDash}         {\mathrel}{AMSb}{"32}
\DeclareMathSymbol{\nVDash}         {\mathrel}{AMSb}{"33}
\DeclareMathSymbol{\ntrianglerighteq}{\mathrel}{AMSb}{"34}
\DeclareMathSymbol{\ntrianglelefteq}{\mathrel}{AMSb}{"35}
\DeclareMathSymbol{\ntriangleleft}  {\mathrel}{AMSb}{"36}
\DeclareMathSymbol{\ntriangleright} {\mathrel}{AMSb}{"37}
\DeclareMathSymbol{\nleftarrow}     {\mathrel}{AMSb}{"38}
\DeclareMathSymbol{\nrightarrow}    {\mathrel}{AMSb}{"39}
\DeclareMathSymbol{\nLeftarrow}     {\mathrel}{AMSb}{"3A}
\DeclareMathSymbol{\nRightarrow}    {\mathrel}{AMSb}{"3B}
\DeclareMathSymbol{\nLeftrightarrow}{\mathrel}{AMSb}{"3C}
\DeclareMathSymbol{\nleftrightarrow}{\mathrel}{AMSb}{"3D}
\DeclareMathSymbol{\divideontimes}  {\mathbin}{AMSb}{"3E}
\DeclareMathSymbol{\varnothing}     {\mathord}{AMSb}{"3F}
\DeclareMathSymbol{\nexists}        {\mathord}{AMSb}{"40}
\DeclareMathSymbol{\Finv}           {\mathord}{AMSb}{"60}
\DeclareMathSymbol{\Game}           {\mathord}{AMSb}{"61}
\DeclareMathSymbol{\eqsim}          {\mathrel}{AMSb}{"68}
\DeclareMathSymbol{\lessdot}        {\mathbin}{AMSb}{"6C}
\DeclareMathSymbol{\gtrdot}         {\mathbin}{AMSb}{"6D}
\DeclareMathSymbol{\ltimes}         {\mathbin}{AMSb}{"6E}
\DeclareMathSymbol{\rtimes}         {\mathbin}{AMSb}{"6F}
\DeclareMathSymbol{\shortmid}       {\mathrel}{AMSb}{"70}
\DeclareMathSymbol{\shortparallel}  {\mathrel}{AMSb}{"71}
\DeclareMathSymbol{\smallsetminus}  {\mathbin}{AMSb}{"72}
\DeclareMathSymbol{\thicksim}       {\mathrel}{AMSb}{"73}
\DeclareMathSymbol{\thickapprox}    {\mathrel}{AMSb}{"74}
\DeclareMathSymbol{\approxeq}       {\mathrel}{AMSb}{"75}
\DeclareMathSymbol{\succapprox}     {\mathrel}{AMSb}{"76}
\DeclareMathSymbol{\precapprox}     {\mathrel}{AMSb}{"77}
\DeclareMathSymbol{\curvearrowleft} {\mathrel}{AMSb}{"78}
\DeclareMathSymbol{\curvearrowright}{\mathrel}{AMSb}{"79}
\DeclareMathSymbol{\digamma}        {\mathord}{AMSb}{"7A}
\DeclareMathSymbol{\varkappa}       {\mathord}{AMSb}{"7B}
\DeclareMathSymbol{\Bbbk}           {\mathord}{AMSb}{"7C}
\DeclareMathSymbol{\hslash}         {\mathord}{AMSb}{"7D}

\catcode `@=12 

\RequirePackage{amsfonts}
\DeclareMathSymbol{\boxdot}       {\mathbin}{AMSa}{"00}
\DeclareMathSymbol{\boxplus}      {\mathbin}{AMSa}{"01}
\DeclareMathSymbol{\boxtimes}     {\mathbin}{AMSa}{"02}
\DeclareMathSymbol{\square}       {\mathord}{AMSa}{"03}
\DeclareMathSymbol{\blacksquare}  {\mathord}{AMSa}{"04}
\DeclareMathSymbol{\centerdot}    {\mathbin}{AMSa}{"05}
\DeclareMathSymbol{\lozenge}      {\mathord}{AMSa}{"06}
\DeclareMathSymbol{\blacklozenge} {\mathord}{AMSa}{"07}
\DeclareMathSymbol{\circlearrowright}   {\mathrel}{AMSa}{"08}
\DeclareMathSymbol{\circlearrowleft}    {\mathrel}{AMSa}{"09}
\DeclareMathSymbol{\leftrightharpoons}  {\mathrel}{AMSa}{"0B}
\DeclareMathSymbol{\boxminus}     {\mathbin}{AMSa}{"0C}
\DeclareMathSymbol{\Vdash}        {\mathrel}{AMSa}{"0D}
\DeclareMathSymbol{\Vvdash}       {\mathrel}{AMSa}{"0E}
\DeclareMathSymbol{\vDash}        {\mathrel}{AMSa}{"0F}
\DeclareMathSymbol{\twoheadrightarrow}  {\mathrel}{AMSa}{"10}
\DeclareMathSymbol{\twoheadleftarrow}   {\mathrel}{AMSa}{"11}
\DeclareMathSymbol{\leftleftarrows}     {\mathrel}{AMSa}{"12}
\DeclareMathSymbol{\rightrightarrows}   {\mathrel}{AMSa}{"13}
\DeclareMathSymbol{\upuparrows}         {\mathrel}{AMSa}{"14}
\DeclareMathSymbol{\downdownarrows} {\mathrel}{AMSa}{"15}
\DeclareMathSymbol{\upharpoonright} {\mathrel}{AMSa}{"16}
 
\DeclareMathSymbol{\downharpoonright}   {\mathrel}{AMSa}{"17}
\DeclareMathSymbol{\upharpoonleft}  {\mathrel}{AMSa}{"18}
\DeclareMathSymbol{\downharpoonleft}{\mathrel}{AMSa}{"19}
\DeclareMathSymbol{\rightarrowtail} {\mathrel}{AMSa}{"1A}
\DeclareMathSymbol{\leftarrowtail}  {\mathrel}{AMSa}{"1B}
\DeclareMathSymbol{\leftrightarrows}{\mathrel}{AMSa}{"1C}
\DeclareMathSymbol{\rightleftarrows}{\mathrel}{AMSa}{"1D}
\DeclareMathSymbol{\Lsh}            {\mathrel}{AMSa}{"1E}
\DeclareMathSymbol{\Rsh}            {\mathrel}{AMSa}{"1F}
\DeclareMathSymbol{\leftrightsquigarrow}{\mathrel}{AMSa}{"21}
\DeclareMathSymbol{\looparrowleft}  {\mathrel}{AMSa}{"22}
\DeclareMathSymbol{\looparrowright} {\mathrel}{AMSa}{"23}
\DeclareMathSymbol{\circeq}       {\mathrel}{AMSa}{"24}
\DeclareMathSymbol{\succsim}      {\mathrel}{AMSa}{"25}
\DeclareMathSymbol{\gtrsim}       {\mathrel}{AMSa}{"26}
\DeclareMathSymbol{\gtrapprox}    {\mathrel}{AMSa}{"27}
\DeclareMathSymbol{\multimap}     {\mathrel}{AMSa}{"28}
\DeclareMathSymbol{\because}      {\mathrel}{AMSa}{"2A}
\DeclareMathSymbol{\doteqdot}     {\mathrel}{AMSa}{"2B}
 
\DeclareMathSymbol{\triangleq}    {\mathrel}{AMSa}{"2C}
\DeclareMathSymbol{\precsim}      {\mathrel}{AMSa}{"2D}
\DeclareMathSymbol{\lesssim}      {\mathrel}{AMSa}{"2E}
\DeclareMathSymbol{\lessapprox}   {\mathrel}{AMSa}{"2F}
\DeclareMathSymbol{\eqslantless}  {\mathrel}{AMSa}{"30}
\DeclareMathSymbol{\eqslantgtr}   {\mathrel}{AMSa}{"31}
\DeclareMathSymbol{\curlyeqprec}  {\mathrel}{AMSa}{"32}
\DeclareMathSymbol{\curlyeqsucc}  {\mathrel}{AMSa}{"33}
\DeclareMathSymbol{\preccurlyeq}  {\mathrel}{AMSa}{"34}
\DeclareMathSymbol{\leqq}         {\mathrel}{AMSa}{"35}
\DeclareMathSymbol{\leqslant}     {\mathrel}{AMSa}{"36}
\DeclareMathSymbol{\lessgtr}      {\mathrel}{AMSa}{"37}
\DeclareMathSymbol{\backprime}    {\mathord}{AMSa}{"38}
\DeclareMathSymbol{\risingdotseq} {\mathrel}{AMSa}{"3A}
\DeclareMathSymbol{\fallingdotseq}{\mathrel}{AMSa}{"3B}
\DeclareMathSymbol{\succcurlyeq}  {\mathrel}{AMSa}{"3C}
\DeclareMathSymbol{\geqq}         {\mathrel}{AMSa}{"3D}
\DeclareMathSymbol{\geqslant}     {\mathrel}{AMSa}{"3E}
\DeclareMathSymbol{\gtrless}      {\mathrel}{AMSa}{"3F}
\DeclareMathSymbol{\bigstar}    {\mathord}{AMSa}{"46}
\DeclareMathSymbol{\between}    {\mathrel}{AMSa}{"47}
\DeclareMathSymbol{\blacktriangledown}  {\mathord}{AMSa}{"48}
\DeclareMathSymbol{\blacktriangleright} {\mathrel}{AMSa}{"49}
\DeclareMathSymbol{\blacktriangleleft}  {\mathrel}{AMSa}{"4A}
\DeclareMathSymbol{\vartriangle}        {\mathrel}{AMSa}{"4D}
\DeclareMathSymbol{\blacktriangle}      {\mathord}{AMSa}{"4E}
\DeclareMathSymbol{\triangledown}       {\mathord}{AMSa}{"4F}
\DeclareMathSymbol{\eqcirc}       {\mathrel}{AMSa}{"50}
\DeclareMathSymbol{\lesseqgtr}    {\mathrel}{AMSa}{"51}
\DeclareMathSymbol{\gtreqless}    {\mathrel}{AMSa}{"52}
\DeclareMathSymbol{\lesseqqgtr}   {\mathrel}{AMSa}{"53}
\DeclareMathSymbol{\gtreqqless}   {\mathrel}{AMSa}{"54}
\DeclareMathSymbol{\Rrightarrow}  {\mathrel}{AMSa}{"56}
\DeclareMathSymbol{\Lleftarrow}   {\mathrel}{AMSa}{"57}
\DeclareMathSymbol{\veebar}       {\mathbin}{AMSa}{"59}
\DeclareMathSymbol{\barwedge}     {\mathbin}{AMSa}{"5A}
\DeclareMathSymbol{\doublebarwedge} {\mathbin}{AMSa}{"5B}
\DeclareMathSymbol{\measuredangle}  {\mathord}{AMSa}{"5D}
\DeclareMathSymbol{\sphericalangle} {\mathord}{AMSa}{"5E}
\DeclareMathSymbol{\varpropto}    {\mathrel}{AMSa}{"5F}
\DeclareMathSymbol{\smallsmile}   {\mathrel}{AMSa}{"60}
\DeclareMathSymbol{\smallfrown}   {\mathrel}{AMSa}{"61}
\DeclareMathSymbol{\Subset}       {\mathrel}{AMSa}{"62}
\DeclareMathSymbol{\Supset}       {\mathrel}{AMSa}{"63}
\DeclareMathSymbol{\Cup}          {\mathbin}{AMSa}{"64}
 
\DeclareMathSymbol{\Cap}          {\mathbin}{AMSa}{"65}
 
\DeclareMathSymbol{\curlywedge}   {\mathbin}{AMSa}{"66}
\DeclareMathSymbol{\curlyvee}     {\mathbin}{AMSa}{"67}
\DeclareMathSymbol{\leftthreetimes} {\mathbin}{AMSa}{"68}
\DeclareMathSymbol{\rightthreetimes}{\mathbin}{AMSa}{"69}
\DeclareMathSymbol{\subseteqq}    {\mathrel}{AMSa}{"6A}
\DeclareMathSymbol{\supseteqq}    {\mathrel}{AMSa}{"6B}
\DeclareMathSymbol{\bumpeq}       {\mathrel}{AMSa}{"6C}
\DeclareMathSymbol{\Bumpeq}       {\mathrel}{AMSa}{"6D}
\DeclareMathSymbol{\lll}          {\mathrel}{AMSa}{"6E}
 
\DeclareMathSymbol{\ggg}          {\mathrel}{AMSa}{"6F}
 
\DeclareMathSymbol{\circledS}     {\mathord}{AMSa}{"73}
\DeclareMathSymbol{\pitchfork}    {\mathrel}{AMSa}{"74}
\DeclareMathSymbol{\dotplus}      {\mathbin}{AMSa}{"75}
\DeclareMathSymbol{\backsim}      {\mathrel}{AMSa}{"76}
\DeclareMathSymbol{\backsimeq}    {\mathrel}{AMSa}{"77}
\DeclareMathSymbol{\complement}   {\mathord}{AMSa}{"7B}
\DeclareMathSymbol{\intercal}     {\mathbin}{AMSa}{"7C}
\DeclareMathSymbol{\circledcirc}  {\mathbin}{AMSa}{"7D}
\DeclareMathSymbol{\circledast}   {\mathbin}{AMSa}{"7E}
\DeclareMathSymbol{\circleddash}  {\mathbin}{AMSa}{"7F}
\DeclareMathSymbol{\lvertneqq}    {\mathrel}{AMSb}{"00}
\DeclareMathSymbol{\gvertneqq}    {\mathrel}{AMSb}{"01}
\DeclareMathSymbol{\nleq}         {\mathrel}{AMSb}{"02}
\DeclareMathSymbol{\ngeq}         {\mathrel}{AMSb}{"03}
\DeclareMathSymbol{\nless}        {\mathrel}{AMSb}{"04}
\DeclareMathSymbol{\ngtr}         {\mathrel}{AMSb}{"05}
\DeclareMathSymbol{\nprec}        {\mathrel}{AMSb}{"06}
\DeclareMathSymbol{\nsucc}        {\mathrel}{AMSb}{"07}
\DeclareMathSymbol{\lneqq}        {\mathrel}{AMSb}{"08}
\DeclareMathSymbol{\gneqq}        {\mathrel}{AMSb}{"09}
\DeclareMathSymbol{\nleqslant}    {\mathrel}{AMSb}{"0A}
\DeclareMathSymbol{\ngeqslant}    {\mathrel}{AMSb}{"0B}
\DeclareMathSymbol{\lneq}         {\mathrel}{AMSb}{"0C}
\DeclareMathSymbol{\gneq}         {\mathrel}{AMSb}{"0D}
\DeclareMathSymbol{\npreceq}      {\mathrel}{AMSb}{"0E}
\DeclareMathSymbol{\nsucceq}      {\mathrel}{AMSb}{"0F}
\DeclareMathSymbol{\precnsim}     {\mathrel}{AMSb}{"10}
\DeclareMathSymbol{\succnsim}     {\mathrel}{AMSb}{"11}
\DeclareMathSymbol{\lnsim}        {\mathrel}{AMSb}{"12}
\DeclareMathSymbol{\gnsim}        {\mathrel}{AMSb}{"13}
\DeclareMathSymbol{\nleqq}        {\mathrel}{AMSb}{"14}
\DeclareMathSymbol{\ngeqq}        {\mathrel}{AMSb}{"15}
\DeclareMathSymbol{\precneqq}     {\mathrel}{AMSb}{"16}
\DeclareMathSymbol{\succneqq}     {\mathrel}{AMSb}{"17}
\DeclareMathSymbol{\precnapprox}  {\mathrel}{AMSb}{"18}
\DeclareMathSymbol{\succnapprox}  {\mathrel}{AMSb}{"19}
\DeclareMathSymbol{\lnapprox}     {\mathrel}{AMSb}{"1A}
\DeclareMathSymbol{\gnapprox}     {\mathrel}{AMSb}{"1B}
\DeclareMathSymbol{\nsim}         {\mathrel}{AMSb}{"1C}
\DeclareMathSymbol{\ncong}        {\mathrel}{AMSb}{"1D}
\DeclareMathSymbol{\diagup}       {\mathrel}{AMSb}{"1E}
\DeclareMathSymbol{\diagdown}     {\mathrel}{AMSb}{"1F}
\DeclareMathSymbol{\varsubsetneq}   {\mathrel}{AMSb}{"20}
\DeclareMathSymbol{\varsupsetneq}   {\mathrel}{AMSb}{"21}
\DeclareMathSymbol{\nsubseteqq}     {\mathrel}{AMSb}{"22}
\DeclareMathSymbol{\nsupseteqq}     {\mathrel}{AMSb}{"23}
\DeclareMathSymbol{\subsetneqq}     {\mathrel}{AMSb}{"24}
\DeclareMathSymbol{\supsetneqq}     {\mathrel}{AMSb}{"25}
\DeclareMathSymbol{\varsubsetneqq}  {\mathrel}{AMSb}{"26}
\DeclareMathSymbol{\varsupsetneqq}  {\mathrel}{AMSb}{"27}
\DeclareMathSymbol{\subsetneq}      {\mathrel}{AMSb}{"28}
\DeclareMathSymbol{\supsetneq}      {\mathrel}{AMSb}{"29}
\DeclareMathSymbol{\nsubseteq}      {\mathrel}{AMSb}{"2A}
\DeclareMathSymbol{\nsupseteq}      {\mathrel}{AMSb}{"2B}
\DeclareMathSymbol{\nparallel}      {\mathrel}{AMSb}{"2C}
\DeclareMathSymbol{\nmid}           {\mathrel}{AMSb}{"2D}
\DeclareMathSymbol{\nshortmid}      {\mathrel}{AMSb}{"2E}
\DeclareMathSymbol{\nshortparallel} {\mathrel}{AMSb}{"2F}
\DeclareMathSymbol{\nvdash}         {\mathrel}{AMSb}{"30}
\DeclareMathSymbol{\nVdash}         {\mathrel}{AMSb}{"31}
\DeclareMathSymbol{\nvDash}         {\mathrel}{AMSb}{"32}
\DeclareMathSymbol{\nVDash}         {\mathrel}{AMSb}{"33}
\DeclareMathSymbol{\ntrianglerighteq}{\mathrel}{AMSb}{"34}
\DeclareMathSymbol{\ntrianglelefteq}{\mathrel}{AMSb}{"35}
\DeclareMathSymbol{\ntriangleleft}  {\mathrel}{AMSb}{"36}
\DeclareMathSymbol{\ntriangleright} {\mathrel}{AMSb}{"37}
\DeclareMathSymbol{\nleftarrow}     {\mathrel}{AMSb}{"38}
\DeclareMathSymbol{\nrightarrow}    {\mathrel}{AMSb}{"39}
\DeclareMathSymbol{\nLeftarrow}     {\mathrel}{AMSb}{"3A}
\DeclareMathSymbol{\nRightarrow}    {\mathrel}{AMSb}{"3B}
\DeclareMathSymbol{\nLeftrightarrow}{\mathrel}{AMSb}{"3C}
\DeclareMathSymbol{\nleftrightarrow}{\mathrel}{AMSb}{"3D}
\DeclareMathSymbol{\divideontimes}  {\mathbin}{AMSb}{"3E}
\DeclareMathSymbol{\varnothing}     {\mathord}{AMSb}{"3F}
\DeclareMathSymbol{\nexists}        {\mathord}{AMSb}{"40}
\DeclareMathSymbol{\Finv}           {\mathord}{AMSb}{"60}
\DeclareMathSymbol{\Game}           {\mathord}{AMSb}{"61}
\DeclareMathSymbol{\eqsim}          {\mathrel}{AMSb}{"68}
\DeclareMathSymbol{\lessdot}        {\mathbin}{AMSb}{"6C}
\DeclareMathSymbol{\gtrdot}         {\mathbin}{AMSb}{"6D}
\DeclareMathSymbol{\ltimes}         {\mathbin}{AMSb}{"6E}
\DeclareMathSymbol{\rtimes}         {\mathbin}{AMSb}{"6F}
\DeclareMathSymbol{\shortmid}       {\mathrel}{AMSb}{"70}
\DeclareMathSymbol{\shortparallel}  {\mathrel}{AMSb}{"71}
\DeclareMathSymbol{\smallsetminus}  {\mathbin}{AMSb}{"72}
\DeclareMathSymbol{\thicksim}       {\mathrel}{AMSb}{"73}
\DeclareMathSymbol{\thickapprox}    {\mathrel}{AMSb}{"74}
\DeclareMathSymbol{\approxeq}       {\mathrel}{AMSb}{"75}
\DeclareMathSymbol{\succapprox}     {\mathrel}{AMSb}{"76}
\DeclareMathSymbol{\precapprox}     {\mathrel}{AMSb}{"77}
\DeclareMathSymbol{\curvearrowleft} {\mathrel}{AMSb}{"78}
\DeclareMathSymbol{\curvearrowright}{\mathrel}{AMSb}{"79}
\DeclareMathSymbol{\digamma}        {\mathord}{AMSb}{"7A}
\DeclareMathSymbol{\varkappa}       {\mathord}{AMSb}{"7B}
\DeclareMathSymbol{\Bbbk}           {\mathord}{AMSb}{"7C}
\DeclareMathSymbol{\hslash}         {\mathord}{AMSb}{"7D}

\begin{document}

\begin{center}
{\LARGE{\bf On Differential Structure for\\
\vskip 0.3 cm
  Projective Limits of Manifolds}}
\vskip 1.2 cm
{\large{\bf M.C. Abbati, A. Mani\`a}\footnote{E-mail: 
Alessandro.Mania@mi.infn.it}}
\vskip 0.5 cm 
Dipartimento di Fisica,\\
 Universit\`a degli Studi di Milano\\
 and I.N.F.N., Sezione di Milano\\
via Celoria 16, 20133, Milano, Italy
\vskip 1cm 28 January 1998\\

\end{center}

\begin{abstract}
We investigate the differential calculus defined by Ashtekar and 
Lewandowski on projective limits 
of manifolds by means of cylindrical smooth functions and compare it 
with the ${\cal C}^\infty$ calculus proposed by Fr\"ohlicher and 
Kriegl in more general context. For products of connected manifolds, a Boman 
theorem is proved, showing the equivalence of the two calculi in this 
particular case. Several examples of projective limits of manifolds are 
discussed, arising in String Theory and in loop quantization of Gauge 
Theories.
\end{abstract}

\section{Introduction}
\par
In the recent literature in mathematical physics one often encounters 
spaces which are projective limits of manifolds. In the loop quantization of 
Gauge Theories as Quantum Gravity and the 2-d Yang Mills Theory, projective 
families of manifolds are widely used to obtain a compact 
space $\overline {{\cal A}/{\cal G}}$ extending the space of connections
 modulo gauge transformations. This procedure allows one to define a 
 diffeomorphism invariant measure on $\overline {{\cal A}/{\cal G}}$ in order 
to get a  Hilbert representation of Wilson loop observables  (\cite{AshIsh}, 
\cite{AshtekarLewandowski}, \cite{Baez}; for a general reference for Loop
 Quantum Gravity, see also the bibliography in \cite{Rov}).

Another example arises in String Theory. Actually, S.Nag and D.Sullivan 
 considered in \cite{Nag Sull} the projective 
family of all finite sheeted compact unbranched coverings of a 
given closed Riemann surface of genus $g\ge 2$, obtaining a universal object, 
 called the universal hyperbolic solenoid. To this 
projective limit of surfaces  corresponds the 
universal Teichm\"uller space ${\cal T}_\infty$, the inductive limit of the 
 family of Teichm\"uller spaces on each surface. ${\cal T}_\infty$ contains the 
Teichm\"uller spaces of surfaces of every genus $g\ge 2$, so that it 
could simultaneously parametrize complex structures on surfaces of all 
topologies and it has be proposed as a fundamental object for a non 
perturbative quantization of String Theory (\cite{Nag}).

 One can ask whether  projective limits of manifolds admit a suitable 
differentiable structure. Among projective limits of 
manifolds there are manifolds (ordinary or modelled on infinite 
dimensional spaces) and  spaces which are not manifolds. 
Examples of such pathology are compact groups. 
The notion of projective limit was introduced by A. Weil (\cite{Weil})
just to discuss the structure of locally compact groups and Weil itself proved
 that every compact group is the projective limit of 
a family of compact Lie groups. This  does  not longer mean that any compact 
group admits some differential structure. Actually, a projective 
limit of a non trivial family of compact Lie groups cannot be a Lie 
group. What is worse, it is well known that compact groups can have a 
wild topological structure. This example shows that, if one research 
for a differential structure on projective limits of manifolds, one is forced to
 a profound enlargement of the usual notions of differential 
structure, still remaining on  commutative differential calculi.

This problem seems not so evident in the case of the hyperbolic 
solenoid introduced by Nag and Sullivan, since this space is just
 a foliated surface, a well understood differential  structure 
(\cite{Moore}). 
 There are serious physical motivations to introduce a differential 
 calculus and differential operators on projective limits arising in 
loop quantization. These limits  can  be very 
different, so that a general treatment appears to be necessary.
 A solution of the problem has been proposed by Ashtekar and 
Lewandowski in \cite{Ashtekar} by choosing as ring of smooth 
functions the set of  the cylindrical smooth functions. Roughly speaking, 
 on a projective limit of manifolds $M= \varprojlim _{j\in J} M_j$, one 
consider to be apt to differential calculus just the smooth
 functions on some manifold $M_j$ of the family. Cylindrical differential 
 forms, vector fields and other differential objects are consequently defined. 
 In \S 2 we introduce projective limits of manifolds, give a short 
account of Ashtekar-Lewandowski calculus, set up tangent bundles 
and give some simple examples.

In the mathematical literature several attempts to generalize 
differential calculus and the notion of differential manifold can be 
found (\cite{FK}, \cite{Kock},  \cite{Michor}, \cite{Moerdijk}). 
In this paper we compare the calculus proposed by Ashtekar and 
Lewandowski  with the ${\cal C}^\infty$ calculus, developed by Fr\"olicher and 
Kriegl in \cite{FK}, of which we give a short account in \S 3.  The 
${\cal C}^\infty$ calculus assumes as starting point the 
duality between smooth curves and smooth functions expressed by the Boman 
theorem  for ordinary manifolds (\cite{Boman}):

1) for every ordinary manifold $M$ a path $c:{\bf R} \to M$ is smooth 
if and only if $f\circ c$ is smooth for every smooth function $f:M\to 
{\bf R}$;

2) a map $\varphi :M\to N$, where $N$ is a ordinary manifold, is 
smooth if and only if $\varphi \circ c$ is a smooth curve in $N$
for every smooth curve $c$ in $M$.

A ${\cal C}^\infty$ structure on a set $X$ is accordingly defined assigning 
 a suitable set of ``curves" $c:{\bf R} \to X$ and a suitable 
set of ``functions" $f: X \to {\bf R}$ such that $f\circ c \in C^\infty 
({\bf R}, {\bf R})$. The ${\cal C}^\infty$ category contains 
ordinary manifolds and has many nice mathematical properties; in particular, 
it is Cartesian closed and closed with respect to projective limits. 
The ${\cal C}^\infty$ calculus has been  proved fruitful in locally convex 
vector spaces where straight lines assure a richness of curves to get a good 
differential calculus. Besides, the notion of ${\cal C}^\infty$ structure and 
${\cal C}^\infty$ maps revealed useful in Gauge Theory to characterize the 
holonomy maps associated to smooth connections (\cite{Lewandowski}), as 
reported in Example II, \S 3, even if differential calculus is not developed 
in this setting. In the general case  the ${\cal C}^\infty$ 
category appears too large to treat differential calculus, since the extension
 of the class of ${\cal C}^\infty$ functions depends on the richness of curves 
and the theory works, as it stands, only when a balance between curves and 
${\cal C}^\infty$ functions is assured. Otherwise one clash with an excess of 
${\cal C}^\infty$ functions and with the difficulty of defining a good 
differential for every ${\cal C}^\infty$ function.  In a general context some 
additional requirements on the duality between curves and  functions could be 
necessary.

For projective limits of manifolds, cylindrical smooth functions are 
just a generating set for the canonical ${\cal C}^\infty$ structure 
and the class of ${\cal C}^\infty$ functions can be remarkably larger. 
 The ring of cylindrical smooth 
functions appears as the minimal choice of functions to be considered in 
differential calculus, and the ring of ${\cal 
C}^\infty$ functions the maximal one.

There are examples of projective limits of manifolds where ${\cal 
C}^\infty$ calculus works well and defines the natural differential 
calculus: for instance, the spaces ${\bf R}^{\bf N}$ and the manifold 
$J^\infty (M,N)$ of jets of infinite order of maps between two ordinary 
manifolds $M$ and $N$, introduced in \S 2 and discussed in \S 3. 
These spaces are Fr\'echet manifolds, on which the ${\cal C}^\infty$  calculus
 gives the standard differential calculus. Here the choice of cylindrical 
smooth functions appears as a unnecessary, even if not severe, restriction.

The two  calculi agree in the particular case of  products of compact
 connected manifolds. Actually, in \S 4 we prove 
a Boman theorem for cylindrical smooth functions on such products. 

In \S 5 we give a first characterization of cylindrical smooth functions 
in terms of the ${\cal C}^\infty$ structure for projective limits of 
compact connected manifolds. One  cannot expect that  ${\cal C}^\infty$ 
functions are cylindrical in the general case. The most relevant 
obstruction is the occurrence of many path components 
which could cause a plenty of ${\cal C}^\infty$ functions. 
 Some examples of projective limits of compact connected 
manifolds are discussed which support a natural structure of foliated 
manifold, as the hyperbolic solenoid. In these cases the appropriate ring 
of ``smooth" functions lies between the rings of cylindrical smooth functions 
and the ring of ${\cal C}^\infty$ functions. 

Finally, we give a short account of the projective limits  introduced in 
 Gauge Theory  to obtain a compact space $\overline {\cal A}$ 
extending the space of smooth connections ${\cal A}$ (\cite{Ashtekar}). 
 This is the most interesting case of projective limits of manifolds, 
since the projections are highly not standard, so that the relation 
between ${\cal C}^\infty$ and cylindrical smooth functions is 
difficult to establish. One could use suitable projective subfamilies or 
different projective families to obtain a compact extension of ${\cal A}$,
 choosed on the basis of physical and mathematical criteria. 
 We suggest that a natural mathematical requirement to 
select these families could be the 
possibility to obtain a satisfactory version of Boman theorem.

\par
\vskip 0.1 cm 
\section{Projective limits of manifolds}
\vskip 0.1 cm
\par  We  start with some standard facts about projective families and 
projective limits of (Hausdorff) topological 
spaces (see  \cite{Eil Ste} or \cite{Eng}).  
\par
A {\sl projective family} of topological  spaces is a family
 $\{ X_j, \pi _{ij}, J \}$, where the index set $J$ is a  directed 
set,  $X_j$ is a topological space for 
each $j \in J$ and the projections  $\pi _{ij}:X_j \to X_i$, defined 
for every pair $i,j\in J$ with $i\le j$, are continuous maps
such that $\pi _{jj}= id _{X_j}$ and $\pi _{ij}\pi _{jk}= \pi _{ik}$
 for $i\le j\le k$. 
\par
 A  element $\{ x_j \}_{j\in J}$ of the product $\Pi _{j \in J} X_j$ is 
called a {\sl thread} if $\pi _{ij}x_j =x_i$ for $i<j$. The  set 
$X= \varprojlim _{j \in J}X_j$  
of all threads  is
 a closed subset of the product and it is called 
the {\sl limit of the projective family} (or the {\sl projective 
limit}).  The maps $\pi _j : X \to 
X_j\quad \pi _j (\{ x_i \}_{j\in J}):= x_j$, also called projections, are 
continuous and open since a basis of the topology of $X$ consists of the 
subsets $\pi _j ^{-1} (U_j )$, with $U_j$ open in $X_j$. 
\par 
\par  
\vskip 0.1 cm 
Let $\{X_j, \pi _{ij}, J\}$ and $\{X_j ^\prime , \pi _{ij}^\prime , J\}$ be  
projective families of topological spaces, with limit $X$ and $X^\prime$, 
respectively. A family $\{ \phi _j \}_{j\in J}$ 
 of continuous mappings $\phi _j :X_j \to X_j ^\prime$ satisfying the 
coherence condition  
$$
\pi _{ij}^\prime \circ \phi _j = \phi _j \circ \pi _{ij}
\quad \forall j\in j,\quad i\le j
$$
 is said a projective family of mappings. The {\sl limit} of the 
projective family of mappings 
$\{ \phi _j \}_{j\in J}$ is the map $\phi :  X \to X^\prime $ defined by
$$
\phi (\{ x_j \}_{j\in J})= \{ \phi _j ( x_j )\}_{j\in J}~.
$$
 The limit map $\phi$ is continuous and is a homeomorphism whenever 
each $\phi _j$ is a homeomorphism.
\par

Each directed subset $J_0$  of $J$ induces a projective subfamily  
 $\{ X_j , \pi _{ij}, J_0 \}$. If  $X_0$ denotes the limit of the 
induced projective subfamily, the map $\pi _{J_0}: X\to X_0$, 
 defined by $\pi _{J_0}\left( \{x_j\}_{j\in J}\right) =\{ x_j \} _{j \in J_0 }$
, is continuous and open (however $\pi _{J_0}$ may be not surjective). If the 
directed subset $J_0$ is cofinal in $J$, then $X$ and $ X_0$ are homeomorphic. 
 We recall that $J_0\subset J$ is a cofinal subset if for every $j \in J$ 
there exists $j_0 \in J_0$ with $j\le j_0$.  
\par
A projective family $\{ X_j , \pi _{ij}, J\}$ is said {\sl trivial} if 
 the projections $\pi _{ij}$ are homeomorphisms for $j$ belonging to 
some cofinal subset $J_0$. The limit of a trivial family is homeomorphic to
 $X_{j_0}$ for every $j_0 \in J_0$.
\par 
If the index set $J$  admits a countable cofinal subset, the projective 
family $\{ X_j ,\pi _{ij} ,J\}$ is called a projective sequence; in this 
case there exists a cofinal subset which can be identified with ${\bf N}$. 
\par
A projective family is said to be {\sl surjective} if the projections $\pi _j$
 are surjective. This implies that all projections $\pi _{ij}$ are 
onto. A projective family of compact spaces in which the $\pi _{ij}$ 
are surjective maps is surjective. The same property holds for 
projective sequences of (non necessarily compact) spaces. 
\par

The limits of general projective families could be  empty or inherit only few 
topological properties. More regular are limits of surjective families or 
limits of compact spaces: a projective limit of compact spaces is non empty 
and compact.  The  limit of a surjective projective family of connected spaces 
is connected.  Beware however that even limits of surjective projective 
sequences of path connected compact sets could  be not path  connected (see  
Examples III,  IV and  V below).  

\vskip 0.2 cm
 We denote by $Cyl _j (X)$  for $j\in J$ the ring of the  functions
$f:X \to {\bf R}$ of the form $f = \pi _j ^* f_j $, for  a continuous function 
$f_j:X_j \to {\bf R}$  (i.e. $f$ is the pullback of some $f_j\in C(X_j)$). 
The graduated ring $Cyl(X)$ of {\sl cylindrical} 
functions on $X$ is the union $\cup _{j \in J} Cyl _j (X)$. The map 
$\pi _j ^* : C (X_j) \to Cyl (X)~ \quad \pi _j ^* f:= f\circ \pi _j$ is a ring 
homomorphism with range  $Cyl _j (X)$ and is injective if $\pi _j $ is onto. 
Thus for surjective projective families each ring $Cyl _j (X)$ can
 be identified with the ring $C(X_j)$.
\par
 We say that $f:X \to {\bf R}$ is {\sl locally cylindrical} if for each 
$x\in X$ there esists a open neighbourhood $U_x$ of $x$ such that the 
restriction $f_{\upharpoonright U_x}$ agrees with the pullback of some 
$f_j\in C(\pi _j(U_x) )$. Locally cylindrical functions are continuous.
\par
 \vskip 0.2 cm
 Projective limits of ordinary (i.e. finite dimensional paracompact smooth)
 manifolds and their differential properties are the argument 
of this paper. Such limits are often considered in the literature 
 and are topological spaces which in general do not support the structure of 
differential manifold. Here a generalization of ordinary differential calculus 
is introduced  appropriate to these spaces. We start with  a formal definition 
to select a relevant class of projective limits.
 
\begin{definition} A projective family  $\{ M_j , \pi _{ij}, J\}$
 such that 
\par
i) $M_j$ are (ordinary) manifolds,
\par
ii) the projections $\pi _{ij} :M_j \to M_i$ are surjective 
submersions, 
\par \noindent 
will be called a {\bf projective family  of manifolds} and  its
 projective limit $M$ a {\bf projective limit of manifolds}.
\end{definition} 
\par

\vskip 0.1 cm

 To introduce elements of a differential structure on $M$ one can use 
an algebraic method: the starting point is the 
choice of a suitable ring of functions on which vector fields are introduced 
as (suitable) derivations. Using algebraic definitions vector fields, 
differential forms, Lie brackets, Lie derivatives and other differential
 operators can also be defined. This is a procedure widely used also in non
 commutative geometry (\cite{Connes},  \cite{ Manin}) and on 
supermanifolds (\cite{Kastler}).  For a projective limit $M$ of manifolds a 
natural choice appears  to be the (Abelian) ring of {\sl smooth cylindrical 
functions} 
$$
Cyl^\infty (M):= 
\cup _{j \in J} Cyl _j ^\infty (M)
$$
where $Cyl_j ^\infty (M):=\{ \pi _j ^* f_j |f_j \in C^\infty (M_j)\}$ can 
be identified with $C^\infty (M_j)$ if the projective family is 
surjective.   One could also use the ring of all 
{\sl smooth locally cylindrical functions of $M$}, denoted by $Cyl 
_\ell ^\infty (M)$. Of course, for a projective limit $M$ of compact
 manifolds this ring agrees with $Cyl ^\infty (M)$.
 The differential calculus based on $Cyl^\infty (M)$ was proposed by Ashtekar 
and Lewandowski in \cite{Ashtekar}. We shortly discuss this structure.
\par
Even if differential calculus on $M$ can be introduced on the basis of 
purely algebraic definitions it is very natural to start more 
geometrically defining an appropriate ``tangent bundle''.
To the projective family $\{ M_j , \pi _{ij}, J\}$
 we can associate the projective family of manifolds 
$\{ TM_j , T\pi _{ij} ,J\}$,
 whose limit we denote by $TM$. 
  One easily sees that  the limit map  $\tau :TM \to M$
  of the projections $\tau _j :TM_j \to M_j$ is continuous
and onto.
\par 

\par
We refer to $(TM, \tau , M)$ as {\sl the tangent bundle} of $M$. The fiber at 
$x$, the {\sl tangent space at $x$},
 is the  vector space 
$T_xM= \varprojlim _{j \in J} T_{x_j}M_j$ (which is a complete 
nuclear locally convex vector space by  Theorem 7.4 in \cite{Schafer}).  
 Notice however that this ``bundle" does not satisfy the local 
triviality condition.
\par

 The tangent bundle $TM$ plays a r\^ole very similar to the tangent 
bundle of a manifold. Actually, for every  $f\in Cyl ^\infty (M)$, 
$f=\pi _j ^* g_j$ the {\sl differential}
$$
{\bf d}f :TM \to {\bf R}\quad  {\bf d}f (v_x):=d_{x_j}g_j 
(v_{x_j})= (T\pi ^* _j dg_j)(v_x)
$$
is well defined, since the differential 
$ d_{x_j}g_j (v_{x_j})$ does not depend on the representation $f=\pi 
^* _j g_j$. One easily recognizes that ${\bf d}f \in Cyl^\infty _j (TM)$ 
whenever $f \in Cyl^\infty _j (M)$ and its restriction ${\bf d}_xf$ on 
the fibre $T_x M$ is continuous and linear. 
 Every  $v_x \in T_x M$ defines a grade preserving derivation at $x$ 
 on $Cyl^\infty (M)$ by 
$$
f \leadsto L_{v_x} f:= {\bf d}f(v_x) ~.
$$
 \par 
We now assume that the projective family of manifolds is surjective and 
 denote by $D_x$ a grade preserving derivation at $x$ defined on 
$Cyl^\infty (M)$ and by $D_{x_j}: C^\infty (M_j)\to C^\infty (M_j)$ the 
induced derivation at $x_j$, for each $j \in J$.
  By finite dimensionality of $T_{x_j}M_j$ there exists a (unique) 
$v_j \in T_{x_j}M_j$ such that $D_{x_j}$ is the Lie derivative $L_{v_j}$. Since 
$D_{x_i} = D_{x_j} \circ \pi _{ij}^*$ for $i\le j$, one easily recognizes 
that the family $\{ v_j \} _{j \in J}$ is a thread. Thus a $v_x \in T_xM$ is 
defined such that $L_{v_x}=D_x$. Therefore the following proposition holds.
\par
\begin{proposition} Let $\{ M_j , \pi _{ij}, J\}$ be 
 a surjective  
projective family of manifolds with limit $M$. For every $x \in M$ 
the tangent space $T_x M$ is isomorphic 
with the space  of all grade preserving derivations at $x$ on  
$Cyl^\infty (M)$.
\end{proposition} 

\noindent
{\bf Remark} The differential ${\bf d}f$ is well defined also if
 $f\in Cyl^\infty _\ell (M)$. Moreover every tangent vector at $x$ defines
 a grade preserving derivation at $x$ on the 
 graded ring $Cyl^\infty _\ell (M)$ and, for surjective projective 
families, $T_x M$ is isomorphic with the space of all grade preserving 
derivations at $x$ of the ring $Cyl^\infty _\ell (M)$.
\par
\vskip 0.3 cm 
It is natural to define  vector fields on $M$ as derivations on $Cyl ^\infty 
(M)$. Given a surjective projective family $\{ M_j ,\pi _{ij}, J\}$, grade
 preserving derivations $D$ on $Cyl ^\infty (M)$ induce on each $M_j$ a 
derivation $D_j$ and the family $\{D_j \}_{j \in J}$ satisfies the 
coherence condition $(\pi _{ij})_* D_j =D_i$ for $i\le j$. 
 The Lie bracket $[D_1, D_2]$ of two grade preserving derivations is the 
derivation associated to the family $\{ [D_{1;j},D_{2;j}] \} _{j\in 
J}$. Thus grade preserving derivations  on $Cyl ^\infty (M)$ form a Lie algebra.

To every grade preserving derivation on $Cyl ^\infty (M)$, a family  
$\{ X_j \} _{j \in J}$
 of vector fields with  $\pi _{ij}^* X_j= X_i$ for $j\le i$ is 
associated and a section 
$X: M\to TM$ is defined by the limit of these vector fields.  
We remark that one can recover the fields $X_j$ by $X$ since $T\pi _j 
\circ X: M\to TM_j$  depends only by the components in $M_j$ and that 
this property characterizes limits of vector fields. 
The set of these limits is a Lie 
algebra with $[X,Y]:= \varprojlim _{j\in J} [X_j,Y_j]$.
Conversely, to every 
limit of  vector fields a grade preserving derivation $D$ on
 $Cyl ^\infty (M)$ is associated (and Lie brackets are conserved).
Thus we get the next proposition.
\begin{proposition} Let $M$ be the limit of a surjective  projective family 
of manifolds $\{ M_j , \pi _{ij}, J\}$. Grade preserving derivations on
 $Cyl ^\infty (M)$ and projective limits 
of vector fields are isomorphic  Lie algebras.
\end{proposition}
 We remark that the objects and the isomorphism in the above proposition 
depend on the given projective family. However, if one take in $J$ a cofinal
 subset $J_0$, the ring of cylindrical functions does not change, 
while one has to consider derivations conserving the grading just for labels 
in $J_0$. 

 One could consider as vector fields on $M$ the limits of vector fields 
arising  by  cofinal subsets $J_0$ of $J$. However this set of fields 
could not admit a Lie bracket. A good Lie bracket is defined if one consider 
 only cofinal subsets of the type $\{ j\in J ~|~ j\ge j_0 \}$ for a given 
$j_0 \in J$, as in \cite{Ashtekar}.
\par
\vskip 0.1 cm
Differential  cylindrical forms are  defined in analogous way as 
cylindrical functions, considering the pullback on $M$ of 
differential forms on the $M_j$. Usual differential operations as 
 Lie derivatives, exterior derivative, etc. and cylindrical cohomology  
are  estabilished. 
\par
\vskip 0.1 cm
Here we introduce same examples of projective limits of manifolds, 
some of which we shall use as toy model in the sequel.

\vskip 0.3 cm 
\noindent {\bf Example I} For a  projective family
 $\{ G_\alpha , \pi _{\alpha \beta}, A\}$, where $G_\alpha$ are  Lie groups
 and the projections are homomorphisms onto, the limit $G$ of is a 
topological group and the projections $\pi _\alpha$ are homomorphisms.
 Notable examples are compact groups: it is well known that every compact
 group is the projective limit of a family of compact Lie groups 
(\cite{Weil}). 
\par
 As Lie groups are parallelizable, the tangent space at the unit  $e$ 
of $G$ is ${\mathfrak g}:=T_e G= \varprojlim _{\alpha \in A} T_{e_\alpha } 
G_\alpha$. Then
$
TG=\varprojlim _{\alpha \in A} TG_\alpha 
=\varprojlim _{\alpha \in A} (G_\alpha \times {\mathfrak g}_\alpha 
)= G\times {\mathfrak g}~.
$
 An exponential map $exp : {\mathfrak g} \to G$ can also be defined as 
the limit of the family of maps $\{ exp _\alpha \} _{\alpha \in A}$. 
This exponential map is continuous, but not open in general. 

Every neighborhood $ U_e$ of the unit $e$ of $G$ contains the kernel 
$H_\alpha$ of the projections $\pi _\alpha$, so that $G$ does admit small 
subgroups  if the normal subgroups $H_\alpha$ are not definitively trivial 
(i.e. if the projective family is not trivial). Therefore the projective 
limit of a not trivial projective family of Lie groups cannot be a ordinary 
 Lie group by the  Yamabe Theorem (\cite{Yamabe}). However a 
projective limit of ordinary Lie groups may be a infinite dimensional Lie 
group. As a simple example we recall that ${\bf R}^{\bf N}$, the space of 
real sequences with the product topology, is a Fr\'echet space, hence an 
Abelian Lie group and it is the projective limit of the Abelian Lie groups 
${\bf R}^d , d \in {\bf N}$.  In this case Yamabe Theorem does not apply: 
${\bf R}^{\bf N}$ admits  indeed small subgroups. 

 We stress that projective limits of a non trivial family of compact 
Lie groups cannot be  Lie groups (as already mentioned in the introduction) 
since in this case the limits are 
compact groups and Yamabe theorem does apply.
\par 
\vskip 0.3 cm
\noindent {\bf Example II}  Now we give an example of  
 a projective sequence of manifolds whose projective limit is  a
 manifold modelled on a  Fr\'echet vector space (\cite{Michor80}). 
\par 
 The set $J^k (M,N)$ of $k$-jets of smooth mappings between manifolds 
 $M$ and $N$, with dimension $m$ and $n$ respectively, is a ordinary
 affine fiber bundle over 
$M\times N$ with fiber at $(x,y)$  the linear space $P^k (m,n):= \prod 
_{j=1}^k L^j _s ({\bf R}^m, {\bf R}^n)$, where
$L^j _s ({\bf R}^m , {\bf R}^n)$ denotes
 the space of $j$-linear symmetric mappings ${\bf 
R}^m \to {\bf R}^n$. 
\par
There are natural projections $\pi _{h,k}: J^k (M,N) \to J^h (M,N)$ 
for $h<k$, 
which in  local charts are truncations of Taylor polynomials up to order 
$h$. 
 As the  projections satisfy the coherence property, the family 
 $\{ J^k (M,N), \pi _{h,k}, {\bf N}\}$ is a projective sequence 
of manifolds (actually, of affine bundles).
\par
The projective limit $J^\infty (M,N)$ of this sequence consists of the 
 Taylor expansions of smooth mappings and  is a manifold modelled on a nuclear 
 Fr\'echet space (\cite{Michor80}). Actually, the limit map 
$J^\infty (M,N) \to M\times N$ of the projective family of projection maps 
is an affine fiber bundle projection with fiber the nuclear 
Fr\'echet space  $P^\infty (m,n)$ of all symmetric formal power series, 
i.e. the product of the spaces $P^k (m,n)$.

 \par 
\vskip 0.3 cm 
\noindent  {\bf Example III} A wide class of projective families of 
manifolds is obtained giving just a manifold $X$ and a map 
 $\phi : X \to X$ which is local diffeomorphism 
onto $X$.  The associated projective sequence is 
$\{ M_n , \pi _{n,m} , {\bf N} \}$ where $M_n= X$ and 
$\pi _{n,m} := \phi ^{m-n}$ for $n < m$. 
 Projective limits of this type arise in the theory of 
dynamical systems (\cite{Sullivan}).

\par
 Since at any point  $x_m\in M_m$ the tangent map $T_{x_m}\pi _{m,n}
:T_{x_m}M_m \to T_{x_n}M_n$ is a linear isomorphism, the 
projective sequence of tangent spaces is trivial, so that the 
tangent space at $x\in M$ is $T_x M= \varprojlim_{n\in 
{\bf N}}T_{x_n}M_m\simeq T_{x_1}X$. As every projective sequence of 
manifolds is  surjective, cylindrical maps are identified with 
smooth functions defined on some $M_n$ . 
\par
 A simple but typical example is the {\sl p-adic solenoid} $\Sigma _p, 
p\in {\bf N}$, $p>1$, constructed as above with $X = S^1$ and 
$\phi : S^1 \to S^1 \quad \phi (z):= z^p$ (see \cite{Eil Ste} and 
\cite{Hew Ros}). The projections are group homomorphisms and coverings. 

It is well known   that $\Sigma _p$ is isomorphic with
 the compact Abelian group $({\bf R}\times \Delta 
_p)/{\bf B}$. Here  $\Delta _p$ is the group of $p$-adic integers, 
i.e. of  formal series ${\mathbf x}= x_0+x_1p+...+x_kp^k+...$ where 
the  coefficients are integers satisfying 
the inequalities $0\le x_k<p,~ k=0,1,2,...$ and ${\bf B}$ denotes the
 subgroup generated by the element $(1,{\bf u})$, with 
${\bf u}\in \Delta _p$ defined by $u_k = \delta _{0,k}$ for 
$k =0,1,...$. 
 We recall that $\Delta _p$ is the projective  limit of the sequence 
of discrete groups ${\bf Z}/p^n {\bf Z}$; therefore it is a Cantor 
group, i.e. a uncountable compact Abelian group which is a perfect 
totally disconnected space.

 An isomorphism with $\Sigma _p$ can be constructed as follows.
 Let $\chi _n :{\bf R}\times \Delta _p \to S^1$ be the epimorphism defined by
$$
\chi _n (t,{\mathbf x})= exp ({2\pi i\over p^n} (t-(x_0 
+x_1p+...+x_{n-1}p^{n-1})).
$$
Since $(\chi _m)^{p^{m-n}}=\chi _n$ for $n<m$, the family $\{ \chi _n, 
{\bf N}\}$ is a projective family of maps. Therefore the limit map $\chi$ 
is defined and is a group homomorphism of ${\bf R}\times 
\Delta _p$ onto $\Sigma _p$. The kernel of $\chi$ is  the group ${\bf 
B}$, so that $\chi$ quotients to the wanted isomorphism $\tilde \chi : 
({\bf R}\times \Delta _p ) /{\bf B} \to \Sigma _p$. The p-adic solenoid 
$\Sigma _p$ is a connected compact  group but it is not arcwise 
connected, not ever locally connected. The path components are 
precisely the images of the continuous homomorphism $\eta _{\bf x}  
:{\bf R}\to \Sigma _p$ defined by
 $\eta _{\bf x}(t):= [(t,{\bf x})]$,  with dense image and 
kernel zero. Moreover the projection ${\bf R} \times \Delta _p \to 
\Delta _p$ quotients to a (not continuous) group epimorphism 
$\Sigma _p \to \Delta _p /{\bf u}{\bf Z}$, whose fibers are exactly the path
 components.  Thus there are uncountably many path components, 
 classified by the Cantor group $\Delta _p / {\bf u}{\bf Z}$, each 
dense (see Remarque 1 in \cite{dixmier}). 
\par
\vskip 0.3 cm \noindent 
{\bf Example IV} It is well known that ${\bf R}$ is the universal 
covering of $S^1$ and that $\pi ^1 (S^1)= {\bf Z}$. For every integer 
$p\in {\bf N}$, consider the subgroups $G_p= p{\bf Z}$ of ${\bf Z}$ 
and the manifolds $M_p:={\bf R}/G_p$, all diffeomorphic to $S^1$. If 
on ${\bf N}$ the ordering is given by $p \preceq q$ if $p$ divides $q$ 
 (so that $G_q \triangleleft G_p$), the quotient map $\pi _{qp}: M_q 
\to M_p$ is defined for $p\preceq q$. So we have a projective 
surjective family of finite sheeted coverings of $S^1$, which are group epimorphisms.
 The limit $\Sigma _\infty$ of this family is a 
compact connected Abelian group and projects on $\Sigma _p$ for every 
 $p\in {\bf N}$. Therefore $\Sigma _\infty$ admits uncountable 
many path components, each dense.

\vskip 0.3 cm 
\noindent 
{\bf Example V}  The {\sl universal laminated surfaces} have been introduced 
and studied by  Nag and Sullivan (\cite{Nag Sull}, \cite{Sullivan} and also 
\cite{Biswas}) in their investigations on the system of Teichm\"uller 
spaces of Riemann surfaces of different genera.
 The relevance of these spaces in path integral quantization of non 
perturbative String Theory was discussed in \cite{Nag}.  For a  closed 
(i.e. compact, connected, without border) Riemann surface $X_g$ of genus $g$,
 equipped with a base point $\star$, the authors considered the  set $J_g$ of
 all homotopy classes of finite sheeted unbranched pointed covering maps 
$\alpha :X_\alpha \to X_g$, where $X_\alpha$ is a closed Riemann surface. This 
set is directed  under the partial ordering given by factorization, i.e. 
$\alpha \preceq \beta $ if there is a commuting triangle of covering maps 
$\beta= \alpha \circ \theta$. The ordered set $J_g$ has a minimum 
$\iota$ corresponding to the identity map on $X_g$. 
 To every $\alpha$ a monomorphism $\pi _1 (\alpha ): \pi _1 
(X_\alpha ,\star ) \to \pi _1 (X_g , \star )$ is associated. Thus 
$\alpha \preceq \beta$ if and only if 
  $Im (\pi _1 (\beta))\subset Im (\pi _1 (\alpha ))$.
\par
  If a universal covering $(X,\star )$ over $(X_g ,\star )$ is fixed,
 $\pi _1 (X_g, \star )$ is identified with  the group 
 $G$ (acting on $X$) of the deck transformations of  $X_g$ and
 $Im (\pi _1 (\alpha ))$ with a subgroup  $G_\alpha$.
 Thus $X_g= X/G$ and  a closed Riemann surface $S_\alpha := X/G_\alpha$ is 
constructed for  each $\alpha$. For $\alpha \preceq \beta$ the projection 
$\pi _{\alpha , \beta }:S_\beta \to S_\alpha$ is defined in the obvious 
way, so one has a projective family $\{ S_\alpha , \pi _{\alpha \beta }, 
J_g \}$ of coverings of $X_g$. Utilising only normal subgroups 
 of $G$ would give a cofinal projective subfamily.
\par

 If $g=1$, $X_g$ is a torus, $({\bf C}, \star )$ is a universal 
covering, ${\bf Z}\oplus {\bf Z}$ is the fundamental group and all 
coverings are also tori. The projective limit is called the {\sl universal 
Euclidean lamination} $E_\infty$.
 The projective family of tori defining 
$E_\infty$ consists of the quotients  ${\bf C}/(p{\bf Z} \oplus q{\bf 
Z})$, $p,q \in {\bf N}$. Hence $E_\infty =\Sigma _\infty \times \Sigma _\infty$.
 
Each surface $X_g$ of genus $\ge 2$ has the Poincar\'e 
hyperbolic half-plane as universal cover.
 The  limit $H_\infty$ projects on surfaces of every genus $\ge 2$. It
 is therefore called 
the {\sl universal hyperbolic lamination}. 
\par 
\vskip 0.5 cm

\section{${\cal C}^\infty$-spaces.}
\par 

We present here the class of ${\cal C}^\infty$ spaces introduced by 
Fr\"ohlicher and Kriegl in \cite{FK}. This is a  very large category 
  containing Fr\'echet manifolds and has nice mathematical properties: 
the  set of  all ${\cal C}^\infty$ functions between each pair of 
$\mathcal {C}^\infty$ spaces has a canonical structure of
 $\mathcal {C}^\infty$ space (Cartesian closedness of the category);
 moreover the ${\cal C}^\infty$ category is closed with 
respect to inductive and projective limits. In particular the last 
property makes the 
proposal of Fr\"ohlicher and Kriegl particularly interesting for us. Previous
 attempts to generalize differential calculus according similar ideas, are
 the  differential spaces of Smith and Chen (\cite{smith} and \cite{chen}).
 As a consequence of Boman Theorem, their  approach 
is essentially equivalent to that of ${\cal C}^\infty$ spaces. 
\par
\vskip 0.1 cm
The idea in $\mathcal {C}^\infty$ spaces is to define a differential 
structure on a set $X$ by a family ${\cal  C}$
 of curves $c: {\bf R} \to X$
 and a family ${\cal S}$ of  functions $f:X \to {\bf R}$
 with the property that ${\cal C}$ and ${\cal S}$ 
determine each other  by the  conditions
\par
$$
{\cal S}= \{ f:X\to {\bf R}~|~ f\circ c \in C^\infty ({\bf R}, {\bf 
R})~\forall c\in {\cal C}\}~ 
$$
$$
{\cal C}= \{ c:{\bf R} \to X~|~ f\circ c \in C^\infty ({\bf R}, {\bf 
R})~\forall f \in {\cal S} \}~.
$$
\vskip 0.1 cm 
 The elements of ${\cal C}$ are called  {\sl structure curves} or
 ${\cal C}^\infty$ curves (or 
simply curves), those of ${\cal S}$ the {\sl structure functions} or 
${\cal C}^\infty$ functions. The pair $({\cal C}, {\cal S})$ 
 is called a $\mathcal {C}^\infty$-structure on $X$ and the triple 
$(X,{\cal C},{\cal S})$ is said a $\mathcal {C}^\infty${\sl -space}.
\par
 A set $C$ of curves in $X$ is  {\sl generating} for $({\cal C}, {\cal 
 S})$ if ${\cal S}=\{f:X\to {\bf R} ~|$ $~f\circ c\in C^\infty ({\bf R}, 
{\bf R})~ \forall c\in C\}$. Analogously, a set of  functions $S$ 
 on $X$ is  generating for $({\cal C}, {\cal S})$ if 
${\cal C}= \{ c: {\bf R} \to X~|~ f\circ c \in 
C^\infty ({\bf R}, {\bf R})~
\forall f \in S \}$.
\par 
A $\mathcal{C}^\infty$ map between $\mathcal {C}^\infty$ spaces $(X_1, 
{\cal C}_1 , {\cal S}_1)$ and $(X_2 , {\cal C}_2 , {\cal S}_2 )$ is a 
map $g:X_1 \to X_2$ satisfying one of  the following equivalent 
conditions: 
$$
g\circ c \in {\cal C}_2 \quad \forall c\in {\cal C}_1;
$$
$$
f\circ g \in {\cal S}_1 \quad \forall f\in {\cal S}_2 ;
$$
$$
f\circ g \circ c \in C^\infty ({\bf R}, {\bf R}) \quad 
\forall f \in {\cal S}_2 ,c\in {\cal C}_1~.
$$
The set of
 all ${\cal C}^\infty$ maps from $X_1$ to $X_2$
 is denoted ${\cal C}^\infty (X_1, X_2)$. 
\par 
On  a ordinary manifold $M$  a $\mathcal {C}^\infty$ 
 structure $({\cal C}, {\cal S})$ is given where 
${\cal C} :=C^\infty ({\bf R},M)$ and  ${\cal S} :=C^\infty (M, {\bf R})$. 
  The set of $\mathcal {C}^\infty$ maps between two manifolds 
$M$ and $N$ is precisely  the set  $C^\infty (M,N)$ of the smooth functions.
 This is a consequence of the Boman Theorem (\cite{Boman}). For every 
${\cal C}^\infty$ space $X$, the set ${\cal C}$ of structure 
curves is precisely ${\cal C}^\infty ({\bf R},X)$, while the set 
${\cal S}$ of structure functions is ${\cal C}^\infty (X,{\bf R})$, 
 briefly denoted by ${\cal C}^\infty (X)$. 
\par
A ${\cal C}^\infty$ structure is defined on products $\prod _{t\in T}X_t$, 
 admitting $\prod _{t\in T} {\cal C}_t$ as set of structure curves,
 where  ${\cal C}_t$ denotes the set of structure curves in $X_t$,  for 
$t\in T$. For every pair $X_1, X_2$ of ${\cal C}^\infty$ 
spaces, the set  ${\cal C}^\infty (X_1, X_2)$
 a canonical ${\cal C}^\infty$ structure is given, setting 
$$
{\cal C}^\infty ({\bf R}, {\cal C}^\infty (X_1, X_2 ))
:= {\cal C}^\infty ({\bf R}\times X_1, X_2)~.
$$
 
\par
 For  ${\cal C}^\infty$ spaces $X_1 ,X_2 , X_3$, one gets the  canonical 
isomorphism 
$$
{\cal C}^\infty (X_1 ,{\cal C}^\infty (X_2 , X_3 ))\simeq {\cal 
C}^\infty (X_1 \times X_2 , X_3 )~.
$$ 
This amounts to the Cartesian closedness of the ${\cal C}^\infty$ 
category, which so appears as a general scenario for 
 Cartesian closed categories of spaces supporting a differential 
calculus and containing ordinary manifolds; for the proof of Cartesian 
closedness, see 1.4.3 in \cite{FK}. However, one encounters serious 
difficulties to define a good tangent space and a differential of 
${\cal C}^\infty$ maps, for a quite general ${\cal C}^\infty$ space. Of course, 
one could proceed as in ordinary manifolds to obtain the kinematical tangent 
space according the following definition.
\par
\vskip 0.1 cm 
\begin{definition} 
Two curves $c_1, c_2 $ of a ${\cal C}^\infty$ space $X$ are said to be 
{\sl tangent} at $x\in X$ if $c_1 (0)=c_2 (0)=x$ and 
$$
\dot {(f \circ c_1 )}(0)= \dot {(f\circ c_2 )}(0)\quad \forall f\in 
{\cal S}~.
$$

The equivalence class  $[c]_x$ of $c$ is called the velocity vector of $c$ 
 at $x$. The set of all velocity vector of curves at $x$
 is the kinematical tangent space at $x$, denoted by ${\cal T} _xX$. 
\end{definition}
\par
 In spite of its name, ${\cal T}_x X$ can fail to have the full 
structure of linear space.   A simple example where  ${\cal T}_xX$ 
is not linear is the following. Take $X= X_1 \cup X_2$ where $X_1$ and $X_2$ 
are orthogonal real lines at $0$ in ${\bf R}^2$.  Structure curves in $X$ are
 smooth curves in ${\bf R}^2$ with values in $X$. 
The kinematical tangent space at $0$ is identified with $X$ itself,
 so it is not linear.

Every $f \in {\cal S}$ does admit a {\sl kinematical differential}  at $x$ 
defined by 
$$
{\bf {\delta}}_x f: {\cal T}_x X \to {\bf R}~\quad v_x\leadsto 
{\bf {\delta }}_x f (v_x):=\dot {(f\circ c)}(0) \quad c\in v_x~.
$$ 
On ${\cal T} X$, the disjoint union $\sqcup _{x \in X}{\cal T} _x X$, 
a surjective map $\tau : {\cal T}X \to X$ is defined by $\tau (v_x):= 
x$. If one assumes $\{ {\bf {\delta} }f | f\in {\cal S} \} \cup 
\{ \tau ^* f |f \in {\cal S}\}$ as a generating set of functions for a  
${\cal C}^\infty$ structure on ${\cal T}X$, the map 
$\tau : {\cal T}X\to X$ is a ${\cal C}^\infty$ map. We refer to 
$({\cal T}X, \tau , X)$  as  the {\sl kinematical tangent bundle}. 
 In particular, if $X$ is a ordinary manifold, then ${\cal T}X$ is just 
the usual tangent bundle $TX$ and ${\bf {\delta }}_xf$ the ordinary 
differential.
\par 
\vskip 0.1 cm 
 Even if the kinematical tangent bundle appears a natural object, there 
are some contexts where another tangent space naturally arises: in the 
case of a projective limit of manifolds $M=\varprojlim  _{j\in J} 
M_j$,  one should assure that good ${\cal C}^\infty$ functions admit a 
 differential defined on $TM=\varprojlim _{j\in J} TM _j$. 
 A right balance between ${\cal C}^\infty$ curves and ${\cal 
C}^\infty$ functions appears necessary to obtain good tangent spaces 
and good differentials for ${\cal C}^\infty$ functions. 
 Actually, in \cite{FK} the general theory of ${\cal 
C}^\infty$ spaces is not fully developed. The main of the book 
concerns ${\cal C}^\infty$ calculus  for a particular class of locally convex 
vector spaces, called {\sl convenient} vector spaces by the authors, 
where straight lines assure a richness of curves to get nice differential 
calculus.

 In a locally convex vector space $E$ the structure curve set ${\cal C}$ 
 is the family of infinitely many 
differentiable curves, where a curve $c:{\bf R} \to E$ is 
differentiable if the derivate
 $\dot c(t):= \lim _{h\to 0} (1/h) (c(t+h)-c(t))$  exists for every 
$t\in {\bf R}$ and the map $t\leadsto \dot c(t)$ is continuous.
 The set ${\cal C}$ does not really depend on 
the locally convex topology of  $E$, but only on the system of its bounded 
sets, so that ${\cal C}^\infty$ functions are not necessarily continuous.
 This cannot be avoided in any calculus, if 
 Cartesian closedness is wanted: actually the 
evaluation $E\times E^\prime \to {\bf R},\quad (x,\ell 
)\leadsto \ell (x)\quad x\in E, \ell \in E^\prime $ (the dual 
space) has to be a ${\cal C}^\infty$ 
function but it is jointly continuous if and  only if $E$ is normable. 
  \par
 Every continuous linear functional on $E$ is a ${\cal 
C}^\infty$ function. A separated locally convex vector space $E$ 
 is called a {\sl convenient vector space} whenever its dual space 
$E^\prime$ is a generating set of functions for the ${\cal C}^\infty$ 
structure of $E$. The name refers to the fact that this class of 
spaces is Cartesian closed and supports a good  calculus. 
\par 
The kinematical tangent space at $x\in E$, for a convenient vector space $E$,
 is precisely $E$. For every $f \in {\cal C}^\infty (E)$ the kinematical 
differential ${\bf {\delta}}_x f$ at $x\in E$ is a continuous 
linear map and agrees with the usual differential $d_x f$  defined by 
$$
d_x f(v)= \lim _{t\to 0} (1/t)(f(x+tv)-f(x))~.
$$
 
 Differential calculus in convenient vector spaces is based on the 
following theorem (see Prop. 4.4.9 of \cite{FK}).
\par
\vskip 0.1 cm
\begin{proposition} Let $E$ be a convenient vector space and $f 
\in {\cal C}^\infty (E,{\bf R})$. Then the differential operator
$$
d: {\cal C}^\infty (E,{\bf R}) \to {\cal C}^\infty (E\times E, {\bf 
R}), \quad f \leadsto df
$$
 is linear and ${\cal C}^\infty$.
\end{proposition}
\vskip 0.1 cm 
As a consequence every ${\cal C}^\infty$ function admits iterated 
differentials of any order. 
\par
\vskip 0.1 cm  Fr\'echet spaces are convenient vector spaces and 
the ${\cal C}^\infty$ calculus agrees with the $C^\infty _c$ calculus 
(we refer the reader to  Appendix A, where a version of Boman Theorem 
for Fr\'echet spaces is given). Thus each ${\cal C}^\infty$ function 
$f$ on a Fr\'echet space $E$ is continuous and its differential in 
 the ${\cal C}^\infty$ calculus agrees with the usual differential 
$df$ in the $C_c ^\infty$ calculus.
 \par
\vskip 0.1 cm 

The theory of convenient  infinite dimensional manifolds has be 
approached in \cite{KM}, where  some manifolds suitable for 
Algebraic Topology are discussed and in the book \cite{book} devoted 
to Global Analysis.  A similar, but different, philosophy has be 
assumed by Michor in his pioniering work (\cite{Michor80}).
 If $M$ is a manifold modelled on Fr\'echet spaces with 
$C_c^\infty$ transition functions, the $C^\infty _c$ functions on $M$
 are precisely the ${\cal C}^\infty$ functions in the ${\cal C}^\infty$ 
structure  generated by $C^\infty _c$ curve and the $C^\infty _c$ curves agree 
with the ${\cal C}^\infty$ curves provided the local model admits bump 
functions. This is the case, for instance, of nuclear Fr\'echet spaces 
(\cite{Lloyd}). 
\par
 We are interested to consider ${\cal C}^\infty$ 
spaces which are not manifolds in any sense, as in the following 
examples.  
\vskip 0.1 cm 
\noindent 
{\bf Example I} The main example of ${\cal C}^\infty$ spaces are 
manifolds. But in foliation theory differential objects arise that are not 
manifolds.  For generalities on foliations see \cite{Candel} and \cite{Moore}. 
\par
We recall that a separable, locally compact metrizable space $M$ is 
said a $d$-dimensional foliated space or a lamination if it  admits a
 cover by open subsets $U_i$ 
(the charts) and homeomorphisms
$$
\varphi _i : U_i \to D_i \times T_i
$$
where $D_i$ is open in ${\bf R}^d$ and $T_i$ any metric space.
The overlap maps are required to be of the form
$$
(\varphi _j \circ \varphi _i ^{-1} ) (z,t)= (\lambda _{ji}(z,t),\tau _{ji}(t))
$$
and of class $C^\infty _l$ : this means that $\lambda _{ji}$ is smooth in the
 variable $z$, with all partial derivatives  continuous in both variables.
Sets of the type $\varphi _i ^{-1} (D_i \times \{ t\})$ glue together 
to form $d$-dimensional manifolds, whose connected components are
 called \emph{leaves}.
 
A $C^\infty _l$ calculus is accordling defined: a  map $f:M\to N$ between foliated spaces is said to be 
of class $C^\infty _l$ if it is continuous, takes leaves to  leaves 
and, for every pair of  charts $\varphi$ in $M$ and $\psi$ in $N$, the local 
expression $\psi \circ f \circ \varphi ^{-1}$ is of class $C^\infty _l$.  
 The inclusion of leaves in $M$ can be not a homeomorphism; it 
is  a homeomorphism with respect to the ``leaf topology",
 obtained by putting on the transversal sets $T_i$ the 
discrete topology. The foliated tangent bundle $T_lM$ is defined as
 the disjoint 
union of the tangent bundles of the leaves and admits a natural structure of 
foliated space defined in a obvious way. 
\par
A natural ${\cal C}^\infty$ structure on $M$ arises, assuming ${\cal C}$ 
to be the set $C^\infty _l ({\bf R}, M)$  of all $C^\infty _l$ curves. The 
range of a $C^\infty _l$ curve is contained in a leaf and is a smooth curve in 
this leaf.   Accordling,  ${\cal C}^\infty $ functions are just 
families $\{ f_\ell \}$ of smooth real functions, one for each leaf 
$\ell$. Thus  ${\cal C}^\infty $ functions may not contain informations 
 on the transversal topology and ${\cal C}^\infty (M, {\bf R})$ agrees with 
 $C^\infty _l (M, {\bf R})$  only if the topology on $M$ is
 the leaf topology.
  The kinematical tangent bundle ${\cal T}M$ coincides as ${\cal 
C}^\infty$ space with the foliated 
tangent bundle and every ${\cal C}^\infty$ function admits iterated 
differentials (along the leaves). $C^\infty _l$ maps are precisely the 
${\cal C}^\infty$ functions whose iterated differentials are 
continuous. This result is a trivial extension of Boman Theorem to 
$d$-dimensional  foliated spaces. 
\par  
Examples of foliated spaces  where  $C^\infty _l  (M) \subset {\cal 
C}^\infty (M)$ strictly are the spaces $\Sigma _p$, $\Sigma _\infty$,
 $E_\infty$ and $H_\infty$ introduced in \S 2. Here we shortly give 
their foliated atlases and we refer to \S 2 for notations. 

\par
A two charts foliated  atlas for $\Sigma _p$ is given by restricting 
the quotient map ${\bf R} \times \Delta _p \to \Sigma _p$ respectively to 
$(0,1)\times \Delta _p$ and $(-1/2 ,1/2)\times \Delta _p$.  
 The leaves of $\Sigma _p$ are precisely the images of the 
homomorphisms $\eta _{\bf x}$, so they are dense. Hence $C^\infty _l$ 
functions are  univocally defined by their restriction to any leaf. 
\par
Foliated atlases can be constructed in a general way for the spaces 
$\Sigma _\infty$, $E_\infty$ and $H_\infty$, owing the fact that they 
are limits of covering manifolds. As an example,  we give a foliated 
atlas for the universal hyperbolic 
lamination $H_\infty$. For a pointed Riemann surface 
$(X_g , \star )$ with $g\ge 2$, choose a universal cover $(X, \star )$.
Fix  an open  subset  $U$ of $X_g$ such that  $U$ 
is the image, by the canonical projection $ X \to X_g$,  of a open 
subset  of the form $B .G$, where $B $ is an open disk contained in a fundamental 
domain in $X$ for the action of $G=\pi ^1 (X_g)$, denoted by the  dot. 
 By the coherence condition we see that for each normal covering surface 
$S_\alpha = X /G_\alpha $, the inverse image of $U$ by the projection 
$\pi _{\alpha, \iota} : S_\alpha \to X_g$ is 
 $(B .G)/G_\alpha \simeq B\times G/G_\alpha$. The groups 
$C_\alpha :=G/G_\alpha$ are finite and form a 
projective family of groups, whose limit ${\mathfrak C}$ is a Cantor 
group. Thus the inverse image  $\pi ^{-1}_\iota (U)$ in $H_\infty$  is  
the inverse limit of the family  $\{ B \times C_\alpha \}$, hence is 
homeomorphic to $B \times {\mathfrak C}$; varing $U$, we obtain a 
foliated atlas. Also in this context, there are uncountably 
many path components, the leaves, parametrized by the Cantor set 
 ${\mathfrak C}$,  each  dense. 
Foliated atlases in $\Sigma _\infty$ and $E_\infty$ are obtained in an 
analogous way, by means of the corresponding universal covering space. 
 There are uncountable many leaves, parametrized by a Cantor 
set, each dense.

\par
 In this paper we just consider real differential structure 
on universal laminations $E_\infty$ and $H_\infty$. More 
appropriately,  complex structures have been defined on universal laminations
 in \cite{Nag Sull}, in which each leaf of $H_\infty$ is identified with 
the Poincar\'e hyperbolic half-plane and leaves of $E_\infty$ with the complex 
plane.
 The Teichm\"uller space of $H_\infty$ is a completion of the inductive limit 
of the Teichm\"uller spaces of the surfaces $S_\alpha$. This Teichm\"uller 
space is expected to play a relevant r\^ole in path quantization of 
String Theory.
 
\vskip 0.3 cm \noindent 
{\bf Example II}  Loop groups are relevant objects in the
 context of the loop representation of Yang Mills Theories and Gravitation 
(\cite{Gambini}, \cite{Rovelli}).
 Different notions of loop group are given in literature and not 
all compatible with a Lie group structure. For instance, the loop 
group considered in  \cite{Bartolo} is embedded in a infinite dimensional Lie
 group, the special extended loop group, but it does not contain any non 
trivial one-parameter subgroup.

 An interesting example of ${\cal C}^\infty$ structure has
 be recently proposed for loop groups (see \cite{Barrett} and 
\cite{Lewandowski}).
 Let  $P(B,G)$ a principal bundle
with $G$ a compact connected Lie group and $B$  a connected manifold.
Two principal bundles $P_1(B,G) $ and $P_2( B,G) $ are
said gauge isomorphic if there exists a bundle isomorphism $\varphi
:P_1\rightarrow P_2$ such that $\varphi ( xg) =\varphi (x)g$, for every 
$x\in P_1$ and $g\in G$. We denote by $\mathcal{G}$
the group of gauge automorphisms of $P( B,G) $.
 By a (parametrized) path in $B$  we mean a continuous map 
$\alpha :[0,1]\rightarrow B$ which is piecewise smooth, i.e. the interval
 $[0,1]$ can be decomposed as finite union of subintervals $[s_i,s_{i+1}]$
 on which $\alpha$ is  smooth. A path $\alpha $ is said a 
 loop if $\alpha (0) =\alpha (1)$; the loop 
$s\leadsto \alpha (1-s)$is denoted  $\alpha ^{-1}$.

On the set of loops based on $\star$, a composition is
defined by

\[
\left( \alpha \circ \beta \right) \left( s\right) =\left\{ 
\begin{tabular}{l}
$\alpha \left( 2s\right) ,$ $s\in \left[ 0,1/2\right] $ \\ 
$\beta \left( 2s-1\right) , s \in \left[ 1/2,1\right] $%
\end{tabular}
.\right. 
\]
The main tool in the loop representation of Yang Mills Theories is however 
the loop group $\mathcal{L}_{\star}$ consisting of the equivalence classes 
of loops based on $\star$, with respect to the relation: 
\begin{equation}
\alpha \sim \beta \;\text{ if \ }H_A(\alpha )=H_A\left( \beta \right) ,\text{
}  \label{equiv}
\end{equation}
 for every connection $A$ on $P(B,G)$, see \cite{Lewandowski}. Here 
$H_A(\alpha )$ denotes the holonomy of $A$ along $\alpha $, defined as follows. 
  The parallel transport  along $\alpha $ of the connection $A$ is an 
equivariant automorphism ${\cal P}^A _ \alpha $ of the fiber
 $P_{\star}$ over the point $\star$; if a point $x_0\in P_{\star}$ is fixed, 
this automorphism is identified with the element $H_A\left( \alpha \right) $ 
of the structure group $G$ satisfying ${\cal P}^A _\alpha (x_0)H_A\left( 
\alpha \right) =x_0$.  We recall that $H_A\left( \alpha \circ \beta \right) 
=H_A\left( \alpha \right) H_A\left( \beta \right)$ and 
$H_A\left( \alpha ^{-1} \right) =H_A\left( \alpha \right) ^{-1}$. 
 If $A_1$ and $A_2$ are gauge equivalent, their holonomy maps are gauge 
 equivalent, i.e. there exists $g\in G$ such that 
$H_{A_1} (\alpha )= gH_{A_2} \left(\alpha \right) g^{-1}$ for every loop 
$\alpha $. 

 The set ${\cal L}_{\star}$ 
becomes a group if its product is defined by 
$\left[ \alpha\right] \circ \left[ \beta \right] 
:=\left[ \alpha \circ \beta \right] ;$ the quotient map 
$H_A:{\cal L}_{\star}\rightarrow G$, $H_A(\left[ \alpha \right] )=
H_A\left( \alpha \right) $ is a homomorphism of groups, called
the holonomy map of the connection $A.$ 

A $\mathcal{C}^\infty $-structure on $\mathcal{L}_{\star}$ is 
generated by the set of curves
$$
\{ 
c:{\bf R}\rightarrow {\cal L}_{\star}, \quad c\left( t\right) 
=\left[ \alpha _t\right] \} , 
$$
where $t\leadsto \alpha _t$ is a homotopy of loops, i.e. 
the map 
$$
h:{\bf R}\times \left[ 0,1\right] \rightarrow B , ~h\left(
t,s\right) :=\alpha _t\left( s\right) 
$$
is continuous and there exists a partition $0=s_1<s_2<...<s_k=1$ of the unit
interval such that, for every $i,$%
\[
h:{\bf R}\times \left( s_i,s_{i+1}\right) \rightarrow B 
\]
is smooth. With respect this $\mathcal{C}^\infty $ structure the group 
operations in ${\cal L}_{\star}$ are ${\cal C}^\infty $.

\vspace{0.1in}

This notion of $\mathcal{C}^\infty $ map is essential to characterize
holonomy maps of smooth connections in the space 
$Hom\left( {\cal L}_{\star},G\right) $ of group homomorphisms. 
 The holonomy map $H_A$ associated to a smooth connection $A$ is
 a $\mathcal{C}^\infty $ map: for every curve in $\mathcal{L}_{\star }$, 
the curve ${\bf R}\ni t\leadsto H_A(\alpha _t)\in G$ is smooth 
since it is obtained
(locally) as solution of a vector field on $G$ depending smoothly on the
parameter $t$ (see   II.3 in \cite{Kobayashi}).
 The correspondence $H:~A\leadsto H_A$
was widely studied (see \cite{Lewandowski} and the bibliography therein). We
summarize their results in the next proposition, where by 
$Hom^\infty \left( {\cal L}_{\star},G\right) $ we denote the space
 of $\mathcal{C}^\infty $ homomorphisms of $\mathcal{L}_{\star}$ in $G$. 
 
\begin{proposition}
\label{Lew}The map $H$ defines a one to one correspondence (up to gauge
equivalence) between smooth connections on smooth $G$-principal bundles on 
$B$ and the elements of $Hom^\infty \left( {\cal L}_{\star},G\right) .$
\end{proposition}

\vspace{0.2 cm}

In \cite{Lewandowski} analogous $\mathcal{C}^\infty $ structures are
considered on path bundles and generalized path principal bundles.
\par
\vskip 1 cm 
\section{Products of Manifolds.}
\par 
 Here we  consider a product space $M=\Pi _{t\in T}M_t$
 of ordinary manifolds $M_t$, where the cardinality of the index set $T$ 
is assumed to be $\leq 2^{\aleph _0}$. $M$ is the  limit of the projective 
surjective family of manifolds $\{ M_j, \pi _{ij}, J\}$, where $J$ denotes 
the directed set of all finite subsets $j$ of $T$ and $M_j = 
\prod _{t\in j} M_t$.  For a subset $S\subset T$ we denote $\pi _S$ the
 projection of $M$ onto $\prod _{t\in S} M_t$.

 We recall that  a canonical ${\cal C}^\infty$ structure is given on  $M$,
 where the set $\mathcal{C}$ of structure curves consists of families 
$c=\{ c_t\} _{t\in T}$ of smooth curves $c_t$  in $M_t.$  The ring 
$Cyl ^\infty (M)$ is in general only a generating set of functions: f.i. 
 functions in $Cyl ^\infty _\ell (M)$  are ${\cal C}^\infty$. We will consider  
also  countably cylindrical $\mathcal{C}^\infty $ functions, i.e. functions 
$f =\pi _{T_0}^{*}f_{T_0}$ which are the pullback of a $\mathcal{C}^\infty $
 function $f_{T_0}:\Pi _{t\in T_0}M_t\rightarrow \mathbf{R}$,  for
a countable $T_0\subset T$. Countably cylindrical ${\cal C}^\infty$ 
functions on $M$ which are not cylindrical do exist; we are indebted with
 A.Kriegl for the following example of a locally cylindrical function on 
$\mathbf{R}^{\mathbf{N}}$  and for the next proposition.
\par\noindent 
\vskip 0.3 cm \noindent 
\textbf{Example} Let $h\in C^\infty ( {\bf R},{\bf R})~,~{\rm supp}~
h\subset \left[ -1/2,1/2\right] ,$ $h\left( 0\right) =1;$ the
function $f:\mathbf{R}^{\mathbf{N}}\rightarrow \mathbf{R},$ $f\left(
x\right) :=\sum_{n=0}^\infty h\left( x_0-n\right) x_n$ is 
 $\mathcal{C}^\infty ( \mathbf{R}^{\mathbf{N}}) $, locally
cylindrical but not cylindrical.

\begin{proposition}
\label{Kriegl}Every $f\in \mathcal{C}^\infty ( \mathbf{R}^{\mathbf{N}%
}) $ is locally cylindrical.
\end{proposition}

\TeXButton{Proof}{\proof} By Theorem \ref{boman} in Appendix A, every 
$\mathcal{C}^\infty $ function $f$ on the Fr\'{e}chet space ${\bf R}^{\bf N}$ 
is a $C^\infty _c$ function, hence  
$df:\mathbf{R}^{\mathbf{N}}\times \mathbf{R}^{\mathbf{N}}\rightarrow 
\mathbf{R}$ is continuous.
Let now $U\times V$ be a connected open subset of 
$\mathbf{R}^{\mathbf{N}}\times \mathbf{R}^{\mathbf{N}}$ such that 
$\left| df\left( x,v\right) \right| <1$ for every 
$\left( x,v\right) \in U\times V$.
  We can assume  that $V=\prod _{n \in \mathbf{N}}V_n$ where $V_n$ are open 
subsets of $\mathbf{R}$ which equal $\mathbf{R}$, except for a finite set 
$\mathbf{N}_0$ of indices. 

Using linearity of $df$ in the second variable one proves that 
 $df(x,v)=0$ if $(x,v)\in U\times V$ and $v_n=0$ for  $n\in \mathbf{N}_0$. 
For $x,y\in U$ and  a smooth curve $c \in {\bf R}^{\bf N}$ joining 
$x$ and $y,$ one has 
\[
f\left( y\right) =f\left( x\right) +\int_0^1df\left( c\left( s\right) ,\dot{c%
}\left( s\right) \right) ds\,. 
\]
If $x_n=y_n$ for every $n\in \mathbf{N}_0$, one can choose $c$ in $U$ such 
that $\dot{c}\left( s\right) _n=0$ $\forall n\in \mathbf{N}_0$ to get 
$f\left( x\right) =f\left( y\right) .$\TeXButton{End Proof}{\endproof}

\smallskip\ 
In this case the restriction to cylindrical smooth functions appears 
unnecessary: using a standard notion of derivative in Fr\'echet spaces 
one obtains the wider class $Cyl_\ell ^\infty$ of smooth functions. 

On a product of manifolds $M=\Pi _{t\in T}M_t$  we can construct
analogous examples of  $\mathcal{C}^\infty $ functions
which are  locally cylindrical, but not cylindrical, if at least one of the 
factors $M_t$ is not compact.
 We have even simple examples of $\mathcal{C}^\infty
$ functions which are not continuous, hence not ever locally cylindrical.
Let $M$ be $({\mathbf 2}\times S)^{\mathbf{N}}$ where $S$ is an ordinary 
manifold and ${\mathbf 2}$ denotes the space consisting of two elements. 
So $M={\mathbf 2}^{\mathbf{N}}\times S^{\mathbf{N}}$ and $\mathcal{C}^\infty$ 
curves in $M$ are maps $s\leadsto \{\xi _n\times c_n\left(
 s\right) \}_{n\in \mathbf{N}},$ where $c_n$ is a smooth curve in $S$ for 
every $n.$ Choose any not continuous function $h$ on ${\mathbf 
2}^{\mathbf{N}}$.  The function $f$ on $M$ defined
by $f\left( \xi ,s\right) =h\left( \xi \right) ,\xi \in {\mathbf 2}^{\mathbf{N}
},s\in S $, is $\mathcal{C}^\infty $ but not continuous.

\vskip 0.3 cm
The main result in this section is that $\mathcal{C}%
^\infty $ functions on a reasonable product $M$ of manifolds 
are continuous and locally
cylindrical, hence cylindrical whenever $M$ is compact. 
 First we prove that,  in a product of connected manifolds, every 
$\mathcal{C}^\infty $ function 
 is countably cylindrical. We need some lemmas.

\begin{lemma}
\label{curvelemma}Let $(M,g)$ be an ordinary connected Riemannian manifold, $%
d_g$ the metric distance, $\{x_n\}_{n\in \mathbf{N}}$ a sequence in $M$
converging to $x,$ such that  $n^nd_g\left( x_n,x\right)
\leq \rho $ for some $\rho >0$ and every $n\in \mathbf{N}$. There exists a 
smooth curve $c$  in $M$
such that $c\left( 1/2^n\right) =x_n$ for every $n$ and $c\left( 0\right) =x.
$
\end{lemma}

\TeXButton{Proof}{\proof}The points $x_n$ and $x$ belong definitively, say 
for $n> \bar n$, to a  normal chart $(U,\exp ^{-1})$; we can 
assume that $x=\exp (0)$, $U=\exp B$ where $B$ is an open ball 
in $T_xM$, so small that $d_g(\exp v,x)=\Vert v\Vert , v\in B$ 
(see for instance Theorem 5.7 Ch.VIII in \cite{Lang}). Since the sequence 
$\left\{ v_n \right\} _{n >\overline{n}}, ~ v_n = \exp ^{-1}(x_n)$, satisfies 
$n^n||v_n||\leq \rho $, we can construct a smooth curve $\gamma $ in $B$ 
 with the properties that $\gamma \left( s\right) =0$ for $s\leq 0$, 
$\gamma \left( 1/2^n\right) =v_n$, and that  
$\gamma \left[ 1/2^{n+1},1/2^n\right] $ is the segment between 
$v_{n+1}$ and $v_n$; by construction, $\gamma$ is flat at every $v_n$ (see  
Proposition 2.3.4 in \cite{FK}).
 The curve $s\leadsto c(s)=\exp \gamma (s)$ is well defined 
and satisfies $c\left( 1/2^n\right) =x_n$ for $n> \overline{n}$. 
As for the remaining points, first suppose that $\overline{n}=1$ so that
 only the point $x_1$ does not belong to the curve $c$. Consider any curve
 $c^\prime$ in the interval $[1/4, +\infty)$ with the properties that 
$c^\prime (s)=x_1 $ for $s\ge 1/2$, $c ^\prime (1/4)=x_2$, with $c^\prime$
 flat at $x_2$, and compose the curve $c$ with $c^\prime$. In the general case
 repeat the procedure adding all the remaining points.
\TeXButton{End Proof}{\endproof}
\smallskip\
\begin{lemma}
\label{Lemmacurva}Let $M=\Pi _{n\in \mathbf{N}}M_n$ be a product of
connected manifolds and $\{ x_k\}_{k \in {\mathbf N}}$ a sequence in $M$
 converging to $x$. Then there exists a subsequence $\left\{ x_{k_r}\right\}$
 and a $\mathcal{C}^\infty $ curve $c$ in $M$ such that
 $c\left( 1/2^r\right) =x_{k_r}$ for every 
$ r\in \mathbf{N}$, $c\left( 0\right) =x.$
\end{lemma}

\TeXButton{Proof}{\proof} Choose a metric $g_n$ on every $M_n$ and
put $d_n\left( x_n,y_n\right) =d_{g_n}(x_n,y_n)(1+d_{g_n}(x_n,y_n))^{-1}$, 
 so that $d_n\left( x_n,y_n\right) \leq 1$ for $x_n,y_n\in M_n$.  $M$ is a 
metric space with the distance $d\left( x,y\right) 
:=\sum_{n=1} ^{+\infty}\frac 1{2^n}d_n\left( x_n,y_n\right) $. 
Extract from $\{x_k\}$ a subsequence $\left\{ x_{k_r}\right\} $
 such that $\{r^rd\left( x_{k_r},x\right) \}$ is bounded, so that even 
the sequence  $\{r^rd_{g_n}\left( x_{k_r;n},x_n\right) \}$ is  bounded for 
each $n$. Using the Lemma \ref{curvelemma} construct  a smooth curve 
$c_n:{\bf R}\rightarrow M_n$, with $c_n\left( 1/2^r\right) =x_{k_r,n}$ and 
$c_n\left( 0\right) =x_n$, for every $n$.
 Then define $c\left( s\right)
=\{c_n\left( s\right) \}_{n\in \mathbf{N}}\in M$.\TeXButton{End Proof}
{\endproof}

\begin{lemma}
\label{lemmanumerabile}Let $M=\Pi _{n\in {\bf N}}M_n$ be a product of
connected manifolds. Then every $f\in {\cal C}^\infty \left( M\right) $
is continuous.
\end{lemma}

\TeXButton{Proof}{\proof}As $M$ is metrizable we have only to prove that $f$
is sequentially continuous. Assume that, for some sequence $\{x_k\}$ of 
$M$ converging to $x$, there exists $\varepsilon >0$ such that 
 $\left| f\left( x_{k}\right) -f\left( x\right) \right| \geq 
\varepsilon $ for all $k$; by considering eventually a subsequence,
 construct by Lemma \ref{Lemmacurva} a $\mathcal{C}^\infty $ curve $c$ 
such that $c(1/2^k)=x_k$, $c(0)=x$. Then 
$f\circ c\in C^\infty ({\bf R},{\bf R})$ and 
 $f\left( x_{k}\right) =\left( f\circ c\right) \left( 1/2^k\right) 
 \rightarrow \left( f\circ c\right)(0)=f(x)$, contradicting the 
assumption.
 \TeXButton{End Proof}{\endproof}

\smallskip\

Let now $M=\Pi _{t\in T}M_t$ and $q\in M$. For every subset
 $S\subset T$ we identify $M_S=\Pi _{t\in S}M_t$ with 
$\{x\in M|\,x_t=q_t,\,\forall t\notin S\}$ and, for $x\in M$, we denote 
by $x_S$ the element defined by $(x_S)_t=x_t$ if $t\in S$, 
$\left( x_S\right) _t=q_t$ if $t\in T-S$. Moreover we consider the subset
 $M_0$ of $M$ consisting of the elements $x$ with  support 
$\{t\in T|\,x_t\neq q_t\}$ at most countable.

\begin{lemma}
Let $M=\Pi _{t\in T}M_t$ be a product of connected manifolds. Every $f\in 
{\cal C}^\infty \left( M\right) $ is sequentially continuous on $M_0.$
\end{lemma}

\TeXButton{Proof}{\proof}Let $\{x_k \} \rightarrow x$, with $x_k,x\in 
M_0$; there exists a subset $S\subset T$, $S$ at most countable, containing
  the supports of $x$ and of the $x_k$; the function 
$f\in {\cal C}^\infty \left( M\right)$, if restricted to $M_S$, is a 
$\mathcal{C}^\infty $ function on $M_S$; then we apply the Lemma 
\ref{lemmanumerabile} to get $f(x_k) \rightarrow f(x)$.
\TeXButton{End Proof}{\endproof}

\smallskip\ 

The following theorem is a consequence of Mazur's results on product of
metrizable spaces (\cite{Mazur}).

\begin{theorem}
\label{continuita}Let $M=\Pi _{t\in T}M_t$ be a product of connected
manifolds. Then every $f\in {\cal C}^\infty \left( M\right)$ is
continuous and countably cylindrical.
\end{theorem}

\TeXButton{Proof}{\proof} The restriction of $f$ to $M_0$ is sequentially
continuous. By Theorem II of \cite{Mazur} there exists a countable subset 
$S_f\subset T$ such that $f\left( x\right) =f\left( x_{S_f}\right)$, 
for  $x\in M_0$. We will prove that $f\left( x\right) =f\left( x_{S_f}\right)$,
 for every $x$ in $M$. We identify the space of the subsets of $T$ with 
${\mathbf 2}^T$ endowed with the product topology and we prove that
 $\varphi _x :{\mathbf 2}^T \to {\bf R}$, 
 $\varphi_x\left( S\right):= 
f\left( x_S\right) -f\left( x_{_{S\cap S_f}}\right)$ is sequentially
continuous. Let $S_n\rightarrow S$, so that, for large $n$, 
$(x_{S_n})_t=\left( x_S\right) _t$ holds for every $t\in T$.
 Applying Lemma \ref{curvelemma}, we can construct, for every $t\in T$, 
a smooth curve $c_t:{\mathbf R}\rightarrow M_t$ 
satisfying $c_t\left( 1/2^n\right) =\left( x_{S_n}\right) _t$ and
 $c_t\left( 0\right) =\left( x_S\right) _t$.
 The curve  $c=\{c_t\}_{t\in T}$ is a ${\cal C}^\infty$ curve and satisfies 
$c\left( 1/2^n\right) =x_{S_n}$, $c\left( 0\right) =x_S$. Since
 $f\circ c\in C^\infty \left( \mathbf{R,R}\right)$, we get 
$f\left( x_{S_n}\right) \rightarrow f\left( x_S\right)$,
 proving  that $\varphi _x$ is sequentially continuous. By Theorem III of
 \cite{Mazur} we conclude that $\varphi _x$ is continuous. For every 
finite set $S$, we have $\varphi _x\left( S\right) =0$ and, applying the 
results in \S 1, Example 3 of \cite{Mazur}, we obtain that 
$\varphi _x\left( S\right) =0$, for every subset $S$ of $T$.
\par
 Then $f=\pi _{S_f}^{*}f_{S_f}$, where $f_{S_f}$ is the restriction of $f$ to 
$M_{S_f}$. Continuity of $f$ follows by Lemma \ref{lemmanumerabile}.%
\TeXButton{End Proof}{\endproof}

\vskip 0.3 cm

Now we come to the problem of defining the differential of ${\cal C}^\infty$ 
functions on products of manifolds. In the case of ${\bf R}^ T$ every 
$f\in {\cal C}^\infty ({\bf R}^T)$ admits a differential: as $f$ is countably 
cylindrical,  we are reduced to the case of 
$f\in \mathcal{C}^\infty \left( \mathbf{R}^{\mathbf{N}}\right) $ discussed 
in Proposition \ref{Kriegl}, obtaining  
 ${\cal C}^\infty ({\bf R}^T)=Cyl ^\infty _\ell ({\bf R}^T)$. 
Therefore the differential of $f$ is ${\bf d}f$, as defined in \S 
2, it  is continuous and satisfies the chain rule. Now we shall prove that 
a similar property holds for ${\cal C}^\infty$ functions on products of 
manifolds   $M=\Pi _{t\in T}M_t$. 
 For a curve $c$ in $M$  we put 
$$
\dot{c}\left( s\right) :=\{\dot{c}_t\left( s\right) \}_{t\in T}\in TM.
$$

\begin{theorem}
Let $M=\Pi _{t\in T} M_t$ be a product of connected geodesically complete 
Riemannian manifolds. Then every $f\in {\cal C}^\infty \left( M\right)$ is 
locally cylindrical.
\end{theorem}

\TeXButton{Proof.}{\proof}In a ordinary  complete
Riemannian manifold we denote by $\gamma _{x,v}$ the geodesic 
curve starting from the point $x$ with velocity $v$ and by  $\Phi$ the flow 
of the spray defined by the metric. We recall that 
$\gamma _{x,v}\left( s\right) =\tau \left( \Phi \left( s,v_x \right) 
\right)$,  where $\tau$ is the tangent projection and 
$v_x\equiv (x,v)$, and that 
$\dot{\gamma}_{x,v}\left( s\right) = \Phi \left( s,v_x \right)$.
 By $\Phi (s+h,v_x )=\Phi (h,\Phi \left( s,v_x\right) )$, we have 
\begin{equation}
\gamma _{x,v}\left( s+h\right) =\gamma _{\gamma _{x,v}\left( s\right) ,\dot{%
\gamma}_{x,v}\left( s\right) }\left( h\right)  \label{geod}
\end{equation}
for every $s,h \in {\bf R}$.
 
We come now to $M=\Pi _{t\in T} M_t$. For $x\in M$, $v\in T_xM$ and 
$s\in {\bf R}$, we denote by $\gamma _{x,v}\left( s\right)$  the  product
 $\left\{ \gamma _{x_t,v_t}\left( s\right) \right\} _{t\in T}$,
 where $\gamma _{x_t,v_t}$ are geodesic curves in $M_t$, and  call 
{\sl geodesic  curve} at $x$ with velocity $v$ the curve 
$\gamma _{x,v}:{\bf R}\rightarrow M,\quad s\leadsto \gamma _{x,v} 
(s)$. The geodesic curve $\gamma _{x,v}$ satisfies formula (\ref{geod}).

 We define now $df:TM\rightarrow \mathbf{R}$ by 
\begin{equation} 
df\left( x,v\right) :=\stackrel{\cdot }{(f\circ \gamma _{x,v})}\left( 
0\right)~.
\label{geode}
\end{equation}

Let $s \leadsto (x(s),v(s))$ be a curve in $TM$.  Applying Boman 
Theorem one easily recognizes that the map $\varphi :{\bf R}^2 \to 
{\bf R}, \quad \varphi (s,h) :=f(\gamma _{x(s),v(s)}(h))$ is smooth. 
Therefore the map $s \leadsto {\partial \over {\partial 
h}}\varphi _{(s,0)}= df(x(s),v(s))$ is smooth. This proves that $df$ 
is a ${\cal C}^\infty$ map, so that  $df$ is continuous by 
Theorem \ref{continuita}.

 Using the Hopf-Rinow Theorem to each component, we get that for every 
$x,y\in M$ there exists a geodesic curve $\gamma $ (possibly not unique) 
joining $x$ to $y$, so that 
$$
f\left( y\right) -f\left( x\right) =
\int_0^1\stackrel{\cdot }{(f\circ \gamma )}\left( s\right) ds.
$$
 We prove that 
$\stackrel{\cdot }{(f\circ \gamma )}\left( s\right)
 =df\left( \gamma \left( s\right) ,\dot{\gamma}\left( s\right) \right)$. 
One has indeed by formulae (\ref{geod}) and (\ref{geode}) 
\begin{eqnarray*}
\lim_{h\rightarrow
0}\frac 1h\left( f\left( \gamma \left( s+h\right) \right) -f\left(
\gamma (s)\right) \right) &=&
\lim_{h\rightarrow 0}\frac 1h\left( f\left(
\gamma _{\gamma \left( s\right) ,\dot{\gamma} \left( s\right)
}\left( h\right) \right) -f\left( \gamma \left( s\right) \right)
\right)\\
&=&df\left( \gamma \left( s\right) ,\dot{\gamma}\left(
s\right) \right).\end{eqnarray*}

Then we remark that $df\left( x, rv \right) =r df\left( x,v \right)$ 
for every  $v\in T_x M$ and $r\in {\bf R}$ (one can use simply a 
reparametrization of curves) and, in particular, that $df\left( x,0\right)=0$. 
Therefore the subset $W$  of $TM$ on which $|df\left( x,v\right) |<1$ is an 
open neighborhood of the zero section. 

We fix $x_0\in M$ and construct a open  set $U \subset W$ of the form 
 $U=\Pi_{t\in T}U_t$, with $U_t=TM_t$ except for a finite 
set $T_0$ of indices, and such that $x_0 \in \tau (U)$.
 If $\left( x,v\right) \in U$ with $v_t=0$ for $t\in T_0,$ then also
 $\left( x,r v\right) \in U$ for every 
 $r\in {\bf R}$, and the condition $\left| df\left( x,r v\right) \right|
=\left| r df\left( x,v \right) \right| < 1$  for every $r\in {\bf 
R}$ implies that $df\left( x,v\right) =0$.
 The set $V:= \tau (U)$ is a open neighborhood of $x_0$. If  $x,y\in V$ 
satisfy $x_t=y_t$ for  $t\in T_0$, there exists
 a geodesic curve $\gamma$ in $V$ joining $x$ to $y$ whose
components $\gamma _t$ are constant for $t\in T_0$. Then 
$$
f\left( y\right) -f\left( x\right) =\int_0^1df\left(
\gamma \left( s\right) ,\dot{\gamma }\left( s\right) \right) ds=0.
$$
 This proves that $f$ is cylindrical on $V$, i.e.  
$f_{\upharpoonright V}=\pi _{T_0}^{*}g_{\upharpoonright V}$ with 
$g\in C^\infty \left( \pi_{T_0}V\right)$.
\TeXButton{End Proof}{\endproof}
\smallskip\ %

Every ordinary manifold admits a complete metric (\cite{Nomizu}), so 
we get the following result.

\begin{theorem}\label{Boman} Let $M= \prod _{t\in T}M_t$ be a product of 
connected manifolds. A function $f$ on $M$ belongs to 
${\cal C}^\infty \left( M\right) $ if and only if it belongs to 
$Cyl_\ell ^\infty (M)$. If the factors $M_t$ are also compact, then every 
$f\in {\cal C}^\infty (M)$ is cylindrical.
\end{theorem}

\vskip 0.1 cm  
\noindent 
{\bf Remark}  The above theorem is a version of Boman Theorem characterizing
 locally cylindrical smooth functions on products $M$ of connected manifolds
 and proves that the ${\cal C}^\infty$ calculus introduced by 
Fr\"ohlicher and Kriegl agrees with the differential calculus proposed 
by Ashtekar and Lewandowski in the case of products of compact 
connected manifolds. In particular, the kinematical tangent space ${\cal 
T}M$ agrees with the tangent space $TM$ and, for each 
$f\in {\cal C}^\infty (M)\equiv Cyl ^\infty _\ell (M)$ and $x\in M$,
the kinematical differential ${\bf \delta}_xf$ agrees with 
the differential ${\bf d}_x f$ defined in \S 2, so that  
$$
{\bf \delta}_{c(s)} f(c(s),\dot{c}(s))= \dot {(f\circ c)}(s) \quad 
\forall s \in {\bf R}
$$
for every curve $c$.  Moreover  the ${\cal C}^\infty$ 
functions are continuous and admit iterated differentials.

 When some factor $M_t$ is not connected, the product $M$ is
 not connected. However, Theorem 13 applies to each connected component 
(of $M$), which results a product of connected manifolds. In this 
setting the ${\cal C}^\infty$ functions on $M$ could be 
not continuous, but they are locally cylindrical (hence continuous) on 
each component.
\par

\vskip 0.3 cm
\noindent {\bf Example.} An interesting example of product of compact
 Lie groups has be proposed as
  a compact extension of the group $\mathcal{G}$ of gauge trasformations of a
principal bundle $P(B,G) $ with compact  connected gauge group $G$ in \cite
{Ashtekar}. We recall that $\mathcal{G}$ is the group of smooth sections of
the associated bundle $P\left[ G\right] $ on $B$, whose fiber on $x\in 
B$ is a group $G_x$ isomorphic with $G$. The group $\mathcal{G}$ is naturally
included in $\overline{\cal G}=\Pi _{x\in B}G_x$ by 
$\eta : {\cal G} \to \overline{\mathcal G},~ \eta (g):=\{ g(x)\}_{x\in B}$.
 Assuming the
cardinality of  $B$ ( space or space-time) to be $\leq 2^{\aleph
_0}$ (i.e. assuming the continuum hypothesis) we obtain that every $\mathcal{%
C}^\infty $-function on $\overline{\mathcal{G}}$ is continuous and
cylindrical (Theorem 13).
\par 
The group $\mathcal{G}$ a natural structure of infinite dimensional Lie
 group can be given. If $B$ is compact, ${\cal G}$ is a Lie group 
modelled on a nuclear Fr\'echet space. As remarked in  \S 3, this 
implies that a ${\cal C}^\infty $ structure for ${\cal G}$ is given, 
admitting $C^\infty _c ({\bf R}, {\cal G})$ as structure curves and
$C^\infty _c ({\cal G}, {\bf R})$ as structure functions. 
 
\begin{proposition} Let $P(B,G)$ a principal bundle with $B$ and $G$ 
compact. The inclusion  $ \eta : ~\mathcal{G}\hookrightarrow \overline{%
\mathcal{G}}$ is a $\mathcal{C}^\infty $ continuous (but not 
open) map. Its image is dense.
\end{proposition}
{\bf Proof}. First we prove that $\eta $ is ${\cal C}^\infty$, i.e. that 
 images of curves in $\mathcal {G}$ are curves
 in $\overline{\mathcal{G}}$.
 Let $s \leadsto g(s)$ be a curve in ${\cal G}$. We have to 
prove that for every $x \in B$ the curve $s \leadsto 
(g(s))(x)\in G_x$ is smooth. This is true since the projection $\pi _x$, 
if restricted to $\mathcal {G}$, agrees with the evaluation map  
$ev_x: {\cal G} \to G_x$ which is $C^\infty _c$ by Corollary 11.7 of 
\cite{Michor80}. This also implies that the inclusion is continuous.
\par
To prove density we only  observe that, given a finite set $\{ x_i \} 
\subset B$ and $g_i\in G_{x_i}$, there exists $g\in \mathcal {G}$ with $g_{x_i}=g_i$ 
for every $i$. 
\par
Completeness of ${\cal G}$ implies that every homeomorphic image of 
${\cal G}$ in a topological group is closed, hence the inclusion 
${\cal G} \hookrightarrow \overline{\cal G}$ cannot be open.
\endproof  
\smallskip\ %

 We stress however, that the group 
$\overline{\mathcal{G}}$ is not a compactification of 
$\mathcal{G}$ endowed with the topology of Fr\'echet Lie group.
\par
\vskip 1 cm

\section{Projective limits of manifolds.} 

The category of ${\cal C}^\infty$ 
spaces is closed with respect to projective limits.
 In  particular, the limit $M$ of a projective family of  manifolds 
$\{ M_j, \pi _{ij} , J\}$ admits a canonical
 $\mathcal {C}^\infty$ structure, 
where the set of structure  curves is 
$$
{\cal C}:= 
\{ c :{\bf R} \to 
M~;~\pi _j \circ c \in C^\infty ({\bf R},M_j) ~\forall j \in J \}~.
$$

 This is precisely the set of ${\cal C}^\infty$ curves in $\prod 
_{j\in J}M_j$ laying in $M\subset \prod _{j\in J} M_j$, so that 
$$
Cyl ^\infty (M) \subset Cyl ^\infty _M (\prod _{j\in J} M_j )
 \subset {\cal C}^\infty _M (\prod _{j\in J} M_j ) \subset {\cal 
C}^\infty (M)~,
$$
where ${\cal C}^\infty _M (\prod _{j\in J} M_j)$ denotes the ring of 
 the restrictions to $M$ of ${\cal C}^\infty $ functions on $\prod 
_{j\in J} M_j$; the ring $Cyl ^\infty _M (\prod _{j\in J} M_j )$ is analogously 
defined.

\begin{proposition} Let $M= \varprojlim _{j\in J} M_j$ be a projective 
limit of manifolds. Then
\par
1) $Cyl ^\infty (M)= Cyl ^\infty _M (\prod _{j\in J} M_j )$;
\par
2) if the factors $M_j$ are compact connected manifolds, a ${\cal 
C}^\infty $ function on $M$ is cylindrical if and only if it admits a 
${\cal C}^\infty$ extension to $\prod _{j\in J}M_j$.
\end{proposition}
\begin{proof} 1) Let $f\in Cyl^\infty (M), ~f=\pi _j ^* f_j$ for some 
$f_j \in C^\infty (M_j)$. Define $f^\sharp =p^* _j f_j$ where $p_j: 
\prod _{\iota\in J} M_\iota \to M_j$ is the Cartesian projection. One easily 
checks that $f^\sharp$ is well defined and that $f=f^\sharp \circ 
i_M$, where $i_M : M \to \prod _{j\in J} M_j$ is the canonical 
inclusion. Consider now any smooth cylindrical function $h$ on 
$\prod _{j\in J}M_j$, with $h= p^* _{J_0} h_0$, $ J_0 =\{ j_1 ,...,j_k 
\}$ and $h_0 : \prod _{j\in J_0}M_j \to {\bf R}$ smooth. It is easy to prove that
 $f=i^* _Mh$ is cylindrical.  Actually, choose $\tilde j$ dominating $J_0$ 
 and define $f_{\tilde 
j}: M_{\tilde j} \to {\bf R}$, $f_{\tilde j} (\tilde x)= h_0 (\pi 
_{\tilde j ,j_1} (\tilde x), ..., \pi_{\tilde j,j_k }(\tilde x))$. 
Then  check that $f=\pi ^* _{\tilde j} f_{\tilde j}$.
\par
 2) This is a immediate consequence of 1) and of Theorem 13.
\end{proof}
\smallskip\ %

The ring $Cyl ^\infty (M)$ is a generating set of functions for the 
canonical ${\cal C}^\infty$ structure and appears just a minimal 
choice for the ring of smooth functions.

 The consistence of ${\cal C}^\infty (M)$ for a projective 
limit of manifolds $M$ could be a problem not so easily estabilished 
as in the case of products of manifolds discussed in the above section. 
The main reason is that the paucity of ${\cal C}^\infty$ curves  produces a 
 plenty of ${\cal C}^\infty$ functions. Even if projective limits of compact
 connected manifolds are connected, they could be not path connected, not 
ever locally path connected (see for instance $\Sigma _p$, $\Sigma _\infty$, 
$E_\infty$ and $H_\infty$ discussed in \S \S   2 and 3). If $M$ admits many 
 path components, there exist ${\cal C}^\infty$ functions on $M$ which are not 
continuous, hence not cylindrical.

If $M$ is the limit of compact connected manifolds $M_j$, then every ${\cal 
C}^\infty$ function admitting a ${\cal C}^\infty$ extension to $\prod 
_{j\in J} M_j$ is cylindrical, hence continuous.  One can ask whether each 
continuous ${\cal C}^\infty$ function $f$ on $M$ is cylindrical. Tiesze 
Extension Theorem assures that $f$ admits a continuous extension 
$\tilde f$ to $\prod _{j\in J} M_j$. This extension could be not a 
${\cal C}^\infty$ map, hence one cannot assure that $f$ is 
cylindrical. An example is given on $\Sigma _p$ (see later). Anyways, 
Theorem II of (\cite{Mazur}) assures that $\tilde f$, hence 
$f$, is countably cylindrical. 

Now we briefly discuss tangent space. Obviously,  
 $TM= \varprojlim _{j\in J}TM_j$ 
is a ${\cal C}^\infty$ space  and the 
projection $\tau :TM \to M$  is a ${\cal C}^\infty$ map. 
\par
\par 
As $M$ is a ${\cal C}^\infty$ space, it admits also a kinematical tangent
 space ${\cal T}M$. A good functoriality would require that ${\cal T}M$ 
agrees with $TM$, as in the case 
of products. This condition allows to differentiate every ${\cal 
C}^\infty $ function on $TM$. If ${\cal C}^\infty (M)= 
Cyl ^\infty _\ell (M)$, the tangent spaces agree, 
but this condition is not necessary, as we shall see discussing the 
examples below.
\vskip 0.3 cm 
\noindent
${\mathbf{J^\infty (M,N)}}$. A simple example  of projective limit of a 
surjective family of non compact manifolds is the space $J^\infty (M,N)$ 
introduced in \S 2, Example II, which is a Fr\'echet manifold modelled on 
 a nuclear Fr\'echet space, the product of a sequence of finite 
dimensional vector spaces. By Theorem \ref{boman} in Appendix A, 
the ${\cal C}^\infty$ functions on $J^\infty (M,N)$ are precisely the 
$C^\infty _c$ functions. Each local expression of a $C^\infty _c$ 
function $f$ on $J^\infty (M,N)$ is locally cylindrical by Theorem 
\ref{Boman}, so that $f$ itself is locally cylindrical.
\par
Of course, even in this case the restriction to smooth cylindrical functions 
appears to be unnecessary. 
\vskip 0.3 cm
\noindent 
{\bf The universal laminations.} We return to the spaces  $\Sigma _p$, 
 $\Sigma _\infty$, $E_\infty$ and $H_\infty$ introduced 
in \S 2. These spaces are foliated  spaces and projective limits  of
 manifolds. Accordling,  they admits two canonical ${\cal C}^\infty$ 
structures. Luckly, these ${\cal C}^\infty$ structures agree.
 Let $M$ stand for $\Sigma _p$, $\Sigma _\infty$, $E_\infty$ or $H_\infty$ 
and $\{ M_j, \pi _j , J\}$ for the corresponding projective family of  
manifolds.  We have to prove that $C^\infty _l$ curves in 
$M$ are precisely the paths $c:{\bf R} \to M$ such that all
 $\pi _j \circ c$ are smooth. Let $c$ be a $C^\infty _l$ curve in 
$M$. Then the projection of $c$ in $M_{j_0}$ is smooth, where $j_0$ 
denotes the minimum of $J$, as one can easily prove 
using the foliated atlas given in \S 3. Since each $\pi _{j,j_0}$ is a 
covering of $M_{j_0}$, even the projection of $c$ on $M_j$ is smooth. 
Conversely, let $c= \{ c_j \} _{j\in J}$ be a thread of smooth curves, 
then $c$ is continuous and contained in a leaf, since leaves are the path 
components. Composing $c$ with the foliated charts we get that $c$ is a 
$C^\infty _l$ curve. 
\par
Coming to ${\cal C}^\infty $ functions, we immediately see that
$$
Cyl ^\infty (M ) \subset C^\infty _l (M ) \subset {\cal C}^\infty (M 
)~.
$$
 We recall that the last inclusion is proper (see  Example II, \S 3).
To show that even the first 
inclusion can be proper, define $f: {\bf R}\times \Delta _p \to {\bf R}$ 
by
$$
f(t,{\bf x}):= \sum _{n=1} ^\infty {1\over 2^n} sin ({2\pi \over p^n} 
(t-(x_0+x_1 p+...+x_{n-1}p^{n-1})).
$$
The function $f$ is a uniform limit of linear combinations of characters, 
so it is continuous. One easily recognizes that its quotient map $\tilde f
 :\Sigma _p \to {\bf R}$ is well defined and a $C^\infty _l$ map. However, 
$\tilde f$ is not a cylindrical map.

 Coming to tangent spaces we see that 
$${\cal T}M= T_lM=TM$$
 as ${\cal C}^\infty$ spaces. The first equality was proved in \S 3.
 We have to prove that $TM=T_l M$. For every $x \in M$ we have
 $T_x M =\varprojlim _{j\in J} T_{x_j} M_j \simeq T_x {\cal L}_x $
since $\varprojlim _{j\in J} T_jM_j$ is a trivial limit and ${\cal 
L}_x$ is a covering of every $M_j$. To show that $T M \simeq T_l M$ 
 as ${\cal C}^\infty$ spaces we can use the same arguments we used above to 
prove that ${\cal C}^\infty$ curves and $C^\infty _l$ curves on $M$ 
agree.

In this example the lack of path connecteness yields a huge quantity 
 of ${\cal C}^\infty$ functions. Nevertheless, this excess of ${\cal C}^\infty$ 
 functions does not create serious problems  for differential calculus. 
Actually, each ${\cal C}^\infty$ function is differentiable on $TM$ owing 
the fact that the various notions of tangent space agree.
 We see therefore that ${\cal C}^\infty$ differential calculus can 
work even if the lack of continuity for ${\cal 
C}^\infty$ functions could be an unpleasant aspect.
 In this example the  relevant ring of functions appears to be 
$C^\infty _l (M)$, which lies between $Cyl ^\infty (M)$ and 
${\cal C}^\infty (M)$. 
\vskip 0.3 cm
\noindent
{\bf Projective limits of manifolds in gauge theories.} In the loop
 quantization of 2-d Yang Mills Theories and Loop Quantum  
Gravity the tool of projective limit has be proved 
useful to embed the configuration space ${\cal A}/{\cal G}$ of the 
theory in a compact space $\overline {{\cal A}/{\cal G}}$ on which 
measures are defined suitable for quantization. Here ${\cal A}$ 
denotes the space of principal connections of a principal bundle 
$P(B,G)$, with $G$ a compact connected group and ${\cal G}$ denotes 
the group of gauge transformations. In the literature many proposal of 
this procedure can be found, whose starting point is a suitable
family of multiloops, graphs or lattices, used as index set for a projective 
 family. Here we briefly  discuss the projective limits of manifolds 
 introduced in \cite{ Ashtekar}.

 Let $B$ be a real analytic connected manifold. By a 
 {\sl parametrized edge} we mean a homeomorphism  $e$ from $[0,1]$ into 
 $B$ such that $e_{\upharpoonright (0,1)} \to B$ is an analytic embedding.
 A {\sl unparametrized } {\sl  edge} is an equivalence
 class of parametrized edges with respect to reparametrization
 by analytic  bijections of the interval $[0,1]$. 
The end points of an edge $e$, called the {\sl vertices} of $e$, and the range 
$e^*$ do not depend by such reparametrizations. 
A {\sl graph}  $\gamma $ in  $M$ consists of finitely many 
unparametrized edges  $e_i$, such that  
\par
1) for $e_i \ne e_j$, $e^*_i\cap e^* _j$ is contained in the set of vertices 
 of $e_i$ and $e_j$;
\par
2) every edge of $\gamma$  is at both sides connected with another 
edge. 
\par
The set $L$ of all the graphs in $M$ can be given a partial 
order, where $\gamma _1 \le \gamma _2$ whenever each edge of $\gamma 
_1$ can be expressed as a  composition of edges of $\gamma _2$ and each 
vertex in $\gamma _1$ is a vertex of $\gamma _2$.
Due to analyticity of edges, $L$ is a directed set.
\par
 For every edge $e$, denote by
 $\widehat {\cal G}_e$ the closed normal subgroup of  ${\cal G}$ consisting 
of gauge transformations  acting as the identity  over the vertices 
of $e$. Define a equivalence relation $\sim _e$ on ${\cal A}$ by 
$$
A\sim _e A^\prime \quad \rm{if}\quad A_{\upharpoonright e^*} =A^\prime 
_{\upharpoonright e^*} \quad 
mod \quad \widehat {\cal G}_{e}~.
$$ 
\par
We denote by ${\cal A}_e$ the quotient space and by $\pi _e :{\cal A} 
\to {\cal A}_e$ the canonical projection. It is well known that, for a given
orientation on $e$,  the parallel transport along $e$ defined by a 
connection $A$, denoted ${{\cal P} _e} ^A$, belongs to 
$Eq(P_{e(0)},P_{e(1)})$, the space of $G$ equivariant 
maps from the fiber $P_{e(0)}$ to the fiber $P_{e(1)}$. The parallel 
transport map ${\cal P}_e :{\cal A} \to Eq (P_{e(0)}, P_{e(1)})$ quotients 
to a one-to-one map $\Lambda _e :{\cal A}_e \to Eq (P_{e(0)},P_{e(1)})$. 
 By means of $\Lambda _e$, a  (analytic) manifold 
structure on ${\cal A}_e$ can be given, which does not depend on the 
choosed orientation:  for $x,x^\prime 
\in B$ the space $Eq (P_x, P_{x^\prime})$ is a compact manifold 
diffeomorphic to $G$ and is canonically diffeomorphic to $Eq(P_{x^\prime }
, P_x)$. 
For a graph $\gamma$, one considers the compact connected manifold 
$$
{\cal A}_\gamma := \prod _{e \in \gamma} {\cal A}_e
$$
and the projection $\pi _\gamma :{\cal A} \to {\cal A}_\gamma ,\quad 
\pi _\gamma := \prod _{e\in \gamma } \pi _e$. For $\gamma <\gamma 
^\prime$ a projection $\pi _{\gamma \gamma ^\prime} : {\cal A}_{\gamma 
^\prime} \to {\cal A} _\gamma$ is defined by $\pi _{\gamma ^\prime 
\gamma}\circ \pi _{\gamma ^\prime}= \pi _\gamma$. This gives  a 
projective surjective family of compact connected manifolds whose limit 
$\overline {\cal A}$ is a compact connected space containing ${\cal A}$
 as dense subset. Elements of $\overline {\cal A}$ are called 
generalized connections. Analogous constructions can be given using 
suitably defined oriented edges and oriented graphs.

 The affine space ${\cal A}$ of the smooth 
connections is modelled on a nuclear Fr\'echet space in the case $B$
 is compact (for the case where $B$ is not compact, see 
\cite{noi}). The inclusion of ${\cal A}$ in $\overline {\cal A}$ is 
 ${\cal C}^\infty$ and continuous,  but it is never a homeomorphism nor 
a ${\cal C}^\infty$ diffeomorphism with its image. This holds also for 
the inclusions of the various Sobolev completions of ${\cal A}$ used 
in the literature. In this sense it is not a true compactification.

A projective family of Lie groups $\{ {\cal G}_\gamma, \pi _\gamma, 
L\}$ is also introduced where ${\cal G}_\gamma := {\cal G}/\widehat
{\cal G}_\gamma$, with $\widehat {\cal G}_\gamma:= \cap _{e\in 
\gamma} \widehat {\cal G}_e$ and $\pi _\gamma$ is the canonical 
projection. The projective limit of this projective family of Lie 
groups is precisely the group $\overline {\cal G}$ considered in \S 3.
 The action of ${\cal G}$ on ${\cal A}$ extends to a action of 
$\overline{\cal G}$ on $\overline{\cal A}$. 

In gauge theories the primary object would be 
$\overline {{\cal A}/{\cal G}}$, the limit of the projective family 
 of orbit spaces ${\cal A}_\gamma / {\cal G}_\gamma$, also considered 
in \cite {Ashtekar}. These orbit spaces fail to be genuine manifolds in
 general. However the authors proved that $\overline {{\cal A}/{\cal G}}$ is 
homeomorphic to $\overline {\cal A} / \overline {\cal G}$ so that 
a differential calculus can be defined on $\overline {{\cal A}/{\cal G}}$ 
by means of $\overline {\cal G}$-equivariant cylindrical smooth maps on 
$\overline {\cal A}$. 

The comparison of ${\cal C}^\infty$ functions with cylindrical smooth 
functions on $\overline {\cal A}$ is a delicate problem, due to the 
complexity of the index set and non triviality of projection maps. 
Even the investigation of the path connectness of $\overline {\cal A}$ 
could reveal a not trivial problem. For the Abelian case a general 
method is reported in Appendix B.

One could hope that the projective limit $\overline {\cal A}$ 
shares some features with the universal laminations. Even in this case 
indeed the projective family is obtained taking quotients of the same 
flat space. However the treatment of these limits requires techniques 
beyond the ones developped in this paper. Moreover, the space 
$\overline {\cal A}$ could be too large for the needs of Quantum 
Field Theory. Actually, some projective subfamilies (as lattices) or 
other projective families (based on multiloops or spin networks 
instead of graphs) are used in the literature, to get analogous 
compactifications of $\overline {\cal A}$.
 Physical and mathematical criteria have to be adopted to select a 
convenient compactification. A good mathematical requirement could be 
to dispose of a suitable Boman theorem to get a fine differential 
calculus.

\vskip 1cm
\noindent
{\bf Acknowledgements}
\vskip 0.3 cm
\noindent 
We are indebt with  A.Kriegl for the  example and Proposition \ref{Kriegl} in 
\S 4. We also would to thank A.Cassa and G. Meloni for stimulating 
discussions. 
\vskip 1cm 
\noindent
{\large{\bf Appendix A}}
\vskip 0.3 cm
\noindent 
 It is well known that standard differential calculus works well
 for finite dimensional vector spaces and for Banach spaces and that 
 a lot of inequivalent differential calculi can be given in  general
 locally convex vector spaces. However, nearly all the main notions of infinite 
differentiability agree in Fr\'echet spaces  (\cite{Ave}, 
\cite{Keller})  with the $C_c^\infty$ differentiability defined as follows.
\par
\vskip 0.1 cm 
Let $U\subset E$ be an open subset of a complete locally convex vector 
space. A mapping $f:U \to F$ is said to be $C^1 _c$ on $U$ if the 
following conditions hold:
\par
1) $ lim _{h\to 0} (1/h)(f(x+hy)-f(x))= Df(x)y$ where $Df(x):E \to F$ 
 is a linear map, for $x\in U$, $y\in E$, $h \in {\bf R}$.
\par
2) The map $Df: U\times E \to F$, $(x,y)\leadsto Df(x)y$ is jointly 
continuous.
\par
 The set of these mappings is denoted $C^1 _c (U,F)$. The spaces $C^k 
_c (U,F)$, $k>1$, are defined by recursion, as the set of the maps in 
$C^{k-1} _c (U,F)$ such that $D^{k-1} f: U \times E^{k-1} \to F$ is 
$C^1 _c$. Then $C^\infty _c (U,F):= \cap _{k\ge 1} C^k _c (U,F)$.
\par
\vskip 0.1 cm 
 More results on $C^\infty _c$ calculus can be found in \cite{Keller} or 
\cite{Michor}.
 In Fr\'echet spaces  the $C^\infty _c$ calculus 
agrees even with the ${\cal C}^\infty$ calculus. 
 We give the proof of this statement which one can find in \cite{FK}, entangled with more general results. A similar 
procedure has be adopted to obtain the results in \S 4.  We recall that 
in a Fr\'echet space $E$ the structure curves are precisely the 
$C^\infty _c$ curves. 
\vskip 0.1 cm 
\begin{lemma}\label{conti} Let $E$ be a Fr\'echet space and $f$ a 
${\cal C}^\infty$ function on $E$. Then $f$ is continuous.
\end{lemma}
{\bf Proof.} Suppose $f$ is not continuous at $x$. Then there exists a sequence 
$\{ x_n\}_{n\in {\bf N}}$ converging to $x$ such that 
$| f(x_n )-f(x)|\ge \epsilon$ for some $\epsilon > 0$. Extract from 
$\{ x_n \}_{n\in {\bf N}}$ a subsequence $\{ x_{n_k} \}$ such that 
$\{ k^k d(x, x_{n_k} )\}$ is bounded, where $d$ is a distance on $E$ 
generating the topology of $E$. Appying Lemma 2.3.4 of \cite{FK}, 
construct a curve $c$ in $E$ such that $c(0)=x$ and 
$c(1/2^k )=x_{n_k}$ for every $k$. The assumption $f\in {\cal C}^\infty 
(E,{\bf R})$ would imply $f(x_{n_k}) \to f(x)$, giving a contradiction.
{\endproof}
\vskip 0.1 cm

The following theorem is the Boman Theorem for $C^\infty _c$ calculus 
on Fr\'echet spaces.
\par
\vskip 0.1 cm 
\begin{theorem}\label{boman} Let  $E$ be a Fr\'echet space and 
$f : E \to {\bf R}$. The 
following statements are equivalent:
\par
(1) $f$ is a ${\cal C}^\infty$ function;
\par
(2) $f$ is a $C_c ^\infty$ function.
\end{theorem}

{\bf Proof.} (1) implies (2). For $x,y \in E$  the map $h \leadsto f(x+hy)$
 belongs to $C^\infty ({\bf R}, {\bf R})$.  We shall  prove 
that  the  map $df:E\times E\to {\bf R}$ defined by 
$$
 df(x,y)=lim _{h\to 0}1/h (f(x+hy)-f(x))
$$
 is a ${\cal C}^\infty$ map, jointly  continuous and linear in the second 
variable. 
\par
To get that $df$ is a ${\cal C}^\infty$ map,  we have to prove that
 the map $\varphi : t\leadsto \varphi (t):=df(x(t),y(t))$
 is smooth, for every pair of curves $t\leadsto x(t)$, $t\leadsto 
 y(t)$ on $E$. Consider the ${\cal C}^\infty$ map 
 $\Phi :{\bf R}^2 \to {\bf R}$ defined by $\Phi (t,h)=f(x(t)+hy(t))$.
  The Boman Theorem on ${\bf R}^2$ gives 
 $\Phi \in C^\infty ({\bf R}^2 ,{\bf R})$. Since
$$
{\partial \over \partial h}\Phi (t,h)|_{h=0}= df(x(t),y(t))=\varphi 
(t)~,  
$$
the map $\varphi$ is smooth. Therefore $df$ is a ${\cal C}^\infty$ 
map. 

By Lemma \ref{conti} $df$ is continuous.  Obviously,  $df$ is homogeneous in 
the second variable. Hence it is linear by Proposition 4.4.22 of \cite{FK}.

We have proved that $f \in C^1 _c (E, {\bf R})$ with $df \in {\cal C}^\infty 
(E\times E, {\bf R})$. By recursivity, this proves that $f \in C^\infty 
_c (E, {\bf R})$.
\par
(2) implies (1). For $f \in C^\infty _c (E,{\bf R})$ and every curve 
$c\in C^\infty _c ({\bf R}, E)$, the composition 
$f \circ c \in C^\infty _c ({\bf R}, {\bf R})= 
C^\infty ({\bf R}, {\bf R})$, so that $f$ is a ${\cal C}^\infty$ 
function.\TeXButton{End Proof}{\endproof}

\vskip 1cm 
\noindent
{\large{\bf Appendix B}}
\vskip 0.3 cm  \noindent 
 Here we refer to the last example in \S 5 and investigate the path 
connectedness of $\overline{\cal A}$ for $G= U(1)$. As proved
 in \cite{AshtekarLewandowski}, one can reduce to a trivial 
principal bundle $P= B\times U(1)$ so that ${\cal A} =A^1 (B)$, the 
space of smooth $1$-forms on $B$. The group ${\cal G}=C^\infty (B, U(1))$ 
acts on ${\cal A}$ by translations $A_\mu \leadsto A_\mu  +g^{-1} 
\partial _\mu g$, so that the action defines a homomorphism of the 
Abelian group ${\cal G}$ in the Abelian group ${\cal A}$. Thus also ${\cal 
A}/{\cal G}$ is an Abelian group.

The triviality of $P$ and commutativity of $U(1)$ imply that
 $\overline {\cal G}= U(1)^B$,  that ${\cal 
A}_\gamma$ is canonically isomorphic to $U(1)^{E(\gamma )}$  (where 
$E(\gamma)$ is the number of edges of $\gamma$) and that the projections 
 $\pi _{\gamma , \gamma ^\prime}$ are group homomorphisms. Hence 
$\overline {\cal A}$ is a compact connected Abelian group. 
 Moreover there exists a short exact sequence of 
compact connected Abelian groups
\begin{equation}
0\to U(1) \to \overline {\cal G} \to \overline {\cal A} \to \overline {{\cal 
A}}/\overline {{\cal G}} \to 0~.
\label{esatta}
\end{equation}

We summarize some of the classical results given in \cite{dixmier} 
about path connecteness of compact connected Abelian groups.

\begin{proposition} Let $X$ a compact connected Abelian group. Then the dual 
group $X ^\dagger$ is discrete and torsion free. The following 
conditions are equivalent:
\par
1) $X$ is path connected;
\par
2) $Ext^1 _{\bf Z} (X^\dagger, {\bf Z})=0$;
\par
3) every element of 
$X$ is of the form $e^{i\lambda}$ where $\lambda \in Hom( X^\dagger , 
{\bf R})$.
\par
If $X^\dagger$ is countable the above conditions 
 are  equivalent to the requirement that $X^\dagger$ is free.
\end{proposition}

  For every graph $\gamma$, the dual group $A_\gamma ^\dagger$ of 
${\cal A}_\gamma$ is the free group generated by the edges in $\gamma$, 
provided that to every edge $e_k$ of $\gamma$ the character 
$$
\chi _{e_k} :{\cal A}_\gamma \to U(1), \quad 
\chi _{e_k} (A_\gamma ):= e^{i\int _{e_k} A}
$$
is associated, where $A\in {\cal A}$ is any representative of 
$A_\gamma$. The dual group  $\overline {\cal A}^\dagger$ of 
$\overline{\cal A}$  is the direct limit  of the dual  groups 
${\cal A}_\gamma ^\dagger $. Every character $\chi$ of 
$ \overline {\cal A}$ belongs to some ${\cal A}_\gamma ^\dagger $, so 
that $\chi = \sum _{e_k \in \gamma} n_k \chi _{e_k}$ and for 
$\overline A \in \overline {\cal A}$ we have
$$
\langle \bar A, \chi \rangle = \langle A_\gamma , \chi  \rangle 
= \prod _k {\chi _{e_k} } ^{n_k} (A_\gamma ) ~.
$$
 In particular, if $\overline {A}= A$ is a smooth connection, it verifies 
$$
\langle A, \chi \rangle = e^{i\lambda (\chi )}~,
$$
where $\lambda \in Hom (\overline {\cal A}^\dagger ,{\bf R})$ is defined by 
$\lambda (\chi )=  \sum _{e_k \in \gamma } n_k \int _{e_k} A$.
 Also the examples of generalized connections given in 
\cite{AshtekarLewandowski} are of the form $e^{i\lambda}$ with 
$\lambda \in Hom (\overline {\cal A}^\dagger , {\bf R})$, so one can 
hope that condition 3) is always verified. 

 Utilizing the exact sequence 
$$
0\to \overline {\widehat {\cal G}} \to \overline {\cal A} \to 
\overline {\cal A} /\overline {\cal G} \to 0~,
$$
 where $\overline {\widehat {\cal G}}:= \overline {\cal G} /U(1)$,  
and Prop. 4, \S 5.5 in \cite{Lambek}, we obtain the exact sequence 
$$
Ext _{\bf Z} ^1 ((\overline {\cal A}/ {\overline {\cal G}})^\dagger 
, {\bf Z}) \to Ext _{\bf Z} ^1 (\overline {\cal A}^\dagger , {\bf 
Z}) \to Ext _{\bf Z} ^1 (\overline {\widehat {\cal G}}^\dagger ,{\bf Z}) \to 0~,
$$
where $Ext _{\bf Z} ^1 (\overline {\widehat {\cal G}}^\dagger ,{\bf Z})=0$, 
since the Abelian group $\overline {\widehat {\cal G}}$ is compact and
 path connected. Therefore $Ext _{\bf Z} ^1 ((\overline  {\cal A}
/\overline {\cal G})^\dagger , {\bf Z}) =0$ would imply that 
$Ext _{\bf Z} ^1 (\overline {\cal A}^\dagger , {\bf Z})=0$. 
This proves that $\overline {\cal 
A}$ is path connected if and only if $\overline {\cal A} /\overline {\cal G}$
 is path connected. 



\begin{thebibliography}{99}




\bibitem{noi} M.C.Abbati, R.Cirelli, A.Mani\`a \emph{The orbit space 
of the action of gauge transformation group on connections.} J. Geom. Phys.
 \textbf{6} (1989) 537-57.


\bibitem{Anandan}  J.Anandan \emph{Holonomy groups in gravity and 
gauge fields.} in: \emph{Proc. Conf. Differential Geometric Methods in 
Physics.} (Trieste 1981) eds. G.Denardo and H.D.Doebner, World 
Scientific 1982.

\bibitem{AshIsh} A.Ashtekar, C.J. Isham \emph{Representations of the 
holonomy algebras of gravity and non-Abelian gauge theories.} Class.  
Quantum Grav. {\bf 9} (1992) 1433-1467. 

\bibitem{AshtekarLewandowski} A.Ashtekar, J.Lewandowski 
\emph{Representation Theory of Analytic Holonomy $C^*$-algebras.}, in: 
 \emph{Knots and Quantum Gravity.} J.C.Baez ed., Oxford University Press, 
Oxford 1994.

\bibitem{Ashtekar}  A.Ashtekar, J.Lewandowski \emph{Differential geometry on
the space of connections via graphs and projective limits.} J. Geom. Phys.
 \textbf{17}
(1995) 191-230.

\bibitem{AshtekarRovelli} A.Ashtekar, C.Rovelli \emph{A loop 
representation for the quantum Maxwell field.} Class. Quantum Grav. 
{\bf 9} (1992) 1121-1150.

\bibitem{Ave} V.I.Averbukh, O.G.Smolyanov \emph{The various definitions 
of the derivative in linear topological spaces.} Russian Math. Surveys 
\textbf{23} (1968) 67-113.  

\bibitem {Baez} J.C. Baez \emph{Diffeomorphism-invariant Generalized Measures 
On the Space of Connections Modulo Gauge Transformations.} in: 
\emph{Proceeding of the Conference on Quantum Topology.} (Manhattan, Kansas, 
March 24-28, 1993) L.Crane and, D.Yetter eds.,  213-223, World 
Scientific, Singapore 1994. 

\bibitem{Barrett}  J.W.Barrett \emph{Holonomy and Path Structures in
General Relativity and Yang-Mills Theory.} Int. J. Theor. Phys.
 \textbf{30 }(1991) 1171-1215. 

\bibitem{Bartolo}  C. Di Bartolo, R.Gambini, J.Griego \emph{The extended loop
group: an infinite-dimensional manifold associated with the Loop 
space.} Commun. Math. Phys. {\bf 158} (1993) 217-240.  

\bibitem{Biswas} I.Biswas, S.Nag \emph{Weil-Petersson Geometry and Determinant 
Bundles on Inductive Limits of Moduli Spaces.} in:  \emph{Contemporary Math. 
Series, (Amer. Math. Soc.)} Bers Colloquium, New York 1995.
 
\bibitem{Boman}  J.Boman \emph{Differentiability of a function and of its
compositions with function of one variable.} Math. Scand. 20 (1967), 249-268. 

\bibitem{Candel} A.Candel \emph{Uniformization of surface 
laminations.} Ann. scient, \'Ec. Norm. Sup. $4^e$ s\'erie, {\bf 
26} (1993), 489-516.

\bibitem{chen} K.T. Chen \emph{Iterated path integrals.} Bull. Am. Math. Soc.
 Vol 83, n.5 (1977), 831-876.





\bibitem{Connes} A.Connes \emph{Non-Commutative Geometry.} Academic 
Press, New York, 1994.


\bibitem{dixmier} J.Dixmier \emph{Quelques propri\'et\'es des groupes 
ab\'eliens localement compacts.} Bull. Sci. Math. {\bf 81}, (1957) 38-48.


\bibitem{Eil Ste} S.Eilenberg, N.Steenrod \emph{Foundations of 
Algebraic Topology.} Princeton University Press, Princeton 1952.
 
\bibitem{Eng} R.Engelking \emph{Outline of General Topology.} 
North-Holland, Amsterdam 1968.


\bibitem{FK}  A.Fr\"ohlicher, A.Kriegl \emph{Linear Spaces and 
Differentiation Theory.} J.Wiley \& Sons 1988.


\bibitem{Gambini} R.Gambini, A.Trias \emph{Gauge Dynamics in the 
C-representation.} Nuclear Physics B278 (1986) 436-448.


\bibitem{Hew Ros} E.Hewitt, K.A.Ross \emph{Abstract Harmonic Analysis. 
Vol I} Die Grundlehren der mathematische Wissenschaft, Band {\bf 115}
 Springer-Verlag 1963.

\bibitem{Hor} J.Horv\'ath \emph{Topological Vector Spaces and 
Distributions 1.} Addison-Wesley, Mass. 1966.

\bibitem{Kastler} A. Jadczyk, D.Kastler \emph{Graded Lie-Cartan Pairs 
I.} Reports on Mathematical Physics {\bf 25}, (1987) 1-51 and
\emph{Graded Lie-Cartan Pairs. II. The Fermionic Differential 
Calculus.} Annals of Physics {\bf 179}, (1987) 169-200. 

\bibitem{Keller} H.H.Keller \emph{Differential calculus in locally 
convex spaces.} L.N.M. {\bf 417} Springer-Verlag, Heidelberg  1974.

\bibitem{Kobayashi} S.Kobayashi, K.Nomizu \emph{Foundations of 
Differential Geometry. Vol I} J.Wiley, N.Y. 1963.


\bibitem{Kock} A.Kock \emph{Syntetic Differential Geometry.} London 
Mathematical Society L.N.S. {\bf 51}, Cambridge 1981.




\bibitem{KM} A.Kriegl, P.Michor \emph{Aspects of the theory of 
infinite dimensional manifolds.} Differential Geometry and its 
Applications 1 (1991) 159-176. 


\bibitem{book} A.Kriegl, P.W. Michor  \emph{The convenient Setting of
 Global Analysis.} Americal Mathematical Society Math. Surv. and Mon.  Vol. 53 
 1997.


\bibitem{Lambek} J.Lambek \emph{Lectures on Rings and Modules.} 
Blaisdell, Waltham 1966. 

\bibitem{Lang}  S.Lang \emph{Differential and Riemannian Manifolds.}
G.T.M. Springer-Verlag, New York Berlin Heidelberg 1994.



\bibitem{Lewandowski}  J.Lewandowski \emph{Group of loops, holonomy maps,
path bundle and path connection.} Class. Quantum Grav. \textbf{10} (1993)
 897-904.

\bibitem{Lloyd} J.Lloyd \emph{Smooth Partitions of Unity on Manifolds.} 
 Trans. Amer. Math. Soc. {\bf 197} (1974) 249-259.

\bibitem{Manin} Y.I.Manin \emph{Topics in Noncommutative Geometry.}
 Princeton University Press, Princeton, New Jersey 1991.


\bibitem{Mazur}  S.Mazur  \emph{On Continuous Mapping on Cartesian 
Products.} Fund. Math. \textbf{39 }(1952) 229-238.

\bibitem{Michor80} P.W.Michor \emph{Manifolds of differentiable 
mappings.} Shiva, Orpington 1980.


\bibitem{Michor} P.Michor \emph{A Convenient Setting for Differential 
Geometry and Global Analysis.} Cahiers Top. Geom. Diff. Vol.XXV-1 (1984) 
63-109, Vol.XXV-2 (1984) 113-178.
  
\bibitem{Moore} C.C.Moore, C.Schochet \emph{Global Analysis on 
Foliated Spaces.} MSRIP Springer-Verlag, New York Berlin Heidelberg 1988.


\bibitem{Moerdijk} I.Moerdijk, G.E.Reyes \emph{Models for smooth 
infinitesimals analysis.} Springer-Verlag, New York Berlin Heidelberg 
1991.  

\bibitem{Naglibro} S.Nag \emph{The complex analytic theory of the 
Teichm\"uller spaces.} Wiley-Interscience, New York 1988.

\bibitem{Nag} S.Nag \emph{Mathematics in and out of String Theory.} 
in: Proceeding of the 37th International Taniguchi Symposium: 
\emph{Topology and Teichm\"uller spaces}.  S.Kojima, et. al. eds., 
1-37, World Scient. 1996.

\bibitem{Nag Sull} S.Nag, D.Sullivan \emph{Teichm\"uller Theory and 
the Universal Period Mapping via Quantum Calculus and the $H^{1/2}$ 
Space on the Circle.} Osaka Journal of Math. \textbf{32} (1995) 1-34.

\bibitem{Nomizu}  K. Nomuzu, H.Ozeki \emph{The existence of complete
Riemannian metrics.} Proc. Amer. Math. Soc. \textbf{12} (1961) 889-891. 


\bibitem{Rov} C.Rovelli \emph{Loop Quantum Gravity.} [Preprint  gr-qc 9710008 ]

\bibitem{Rovelli} C.Rovelli, L.Smolin \emph{Loop space representation 
of Quantum General Relativity.} Nuclear Physics B331 (1990) 80-152.

\bibitem{Schafer} H.H.Sch\"afer \emph{Topological Vector Spaces.} 
Macmillan, New York 1966.

\bibitem{smith} J. W. Smith \emph{The de Rahm theorem for general spaces.}
 Tohoku Math J. (2) {\bf 18} (1966), 115-137.

\bibitem{Sullivan} D.Sullivan \emph{Linking the universalities of 
Milnor-Thurston Feigenbaum and Ahlfors-Bers.} in: \emph{Topological 
Methods in Modern Mathematics.}, eds. L.R.Goldberg 
and A.V.Phillips,  Publish or Perish, 1993. 

\bibitem{Weil} A.Weil \emph{L'integration dans les groupes 
topologiques et ses applications.} Hermann, Paris 1940.

\bibitem{Yamabe} H.Yamabe \emph{A generalization of a theorem of 
Gleason.} Ann.Math. \textbf{58} (1953) 351-65.





\end{thebibliography}
\end{document}